\documentclass[a4paper,11pt,superscriptaddres]{article}
\pdfoutput=1 

\usepackage{jheppub} 

\usepackage[T1]{fontenc} 

\usepackage{slashed}
\usepackage{amsmath}
\usepackage{amsfonts}
\usepackage{bbm}
\usepackage{dsfont}
\usepackage{feynmp}
\usepackage{slashed}

\usepackage{longtable}
\usepackage{multirow}

\usepackage{hhline}
\usepackage{rotating}
\usepackage{array}
\usepackage{float}
\usepackage{afterpage}

\usepackage[version=4]{mhchem}

\usepackage[dvipsnames]{xcolor}

\usepackage[lofdepth,lotdepth,caption=false]{subfig}
\usepackage{graphicx}
  {}

\allowdisplaybreaks

\def\bseq{\begin{subequations}}
\def\eseq#1{\label{#1}\end{subequations}}
\def\eseqn{\end{subequations}}

\def\identity{\mathds{1}}

\def\vect#1{ \mathbf{#1} }
\def\bra#1{ \langle #1 | }
\def\ket#1{ | #1 \rangle }

\preprint{CALT-TH-2021-018}

\title{Extended Calculation of Dark Matter-Electron Scattering in Crystal Targets}

\author[a]{Sin\'{e}ad~M.~Griffin,}
\author[a]{Katherine~Inzani,}
\author[b]{Tanner~Trickle,}
\author[b]{Zhengkang~Zhang}
\author[b]{and Kathryn~M.~Zurek}

\affiliation[a]{Material Sciences Division and Molecular Foundry,\\ Lawrence Berkeley National Laboratory, Berkeley, CA 94720, USA}
\affiliation[b]{Walter Burke Institute for Theoretical Physics,\\ California Institute of Technology, Pasadena, CA 91125, USA}

\abstract{
    We extend the calculation of dark matter direct detection rates via electronic transitions in general dielectric crystal targets, combining state-of-the-art density functional theory calculations of electronic band structures and wave functions near the band gap, with semi-analytic approximations to include additional states farther away from the band gap. 
	We show, in particular, the importance of all-electron reconstruction for recovering large momentum components of electronic wave functions, which, together with the inclusion of additional states, has a significant impact on direct detection rates, especially for heavy mediator models and at $\mathcal{O}(10\,\text{eV})$ and higher energy depositions. 
	Applying our framework to silicon and germanium (that have been established already as sensitive dark matter detectors), we find that our extended calculations can appreciably change the detection prospects. 
	Our calculational framework is implemented in an open-source program \texttt{EXCEED-DM} (EXtended Calculation of Electronic Excitations for Direct detection of Dark Matter), to be released in an upcoming publication.
}

\begin{document} 
\maketitle
\flushbottom

\section{Introduction}
\label{sec:intro}

\noindent
Electronic excitations have been established as an alternative to nuclear recoils in direct detection of sub-GeV dark matter (DM). Nuclear recoil searches lose sensitivity at lower DM masses due to kinematic mismatch between the DM and heavier nuclei, whereas electronic transitions can potentially extract all of the DM kinetic energy during a DM-electron scattering event by excitation across an energy gap. Proposed targets, including noble gas atoms with ${\cal O}(10\, \rm eV)$ ionization energies~\cite{Essig:2011nj,Graham:2012su,Lee:2015qva,Essig:2017kqs,Catena:2019gfa,Agnes:2018oej,Aprile:2019xxb,Aprile:2020tmw},
semiconductors with ${\cal O}(\rm eV)$ electronic band gaps~\cite{Essig:2011nj,Graham:2012su,Essig:2012yx,Lee:2015qva,Essig:2015cda,Derenzo:2016fse,Hochberg:2016sqx,Bloch:2016sjj,Kurinsky:2019pgb,Trickle:2019nya, Griffin:2019mvc,Griffin:2020lgd,Du:2020ldo}, and superconductors and Dirac materials with ${\cal O}(\rm meV)$ band gaps~\cite{Hochberg:2015pha,Hochberg:2015fth,Hochberg:2016ajh,Hochberg:2017wce,Coskuner:2019odd,Geilhufe:2019ndy,inzani2021prediction}, extend the reach on DM mass well below the limit of nuclear recoil. Experimental searches using dielectric crystal targets are currently underway, specifically with Si (DAMIC~\cite{deMelloNeto:2015mca,Aguilar-Arevalo:2019wdi,Settimo:2020cbq}, SENSEI~\cite{Tiffenberg:2017aac,Crisler:2018gci,Abramoff:2019dfb,Barak:2020fql}, SuperCDMS~\cite{Agnese:2014aze, Agnese:2015nto, Agnese:2016cpb, Agnese:2017jvy, Agnese:2018col, Agnese:2018gze, Amaral:2020ryn}) and Ge (EDELWEISS~\cite{Armengaud:2018cuy, Armengaud:2019kfj, Arnaud:2020svb}, as well as SuperCDMS) which have been predicted to have excellent sensitivity down to $\mathcal{O}(\text{MeV})$ DM masses based on their $\mathcal{O}(\text{eV})$ band gaps.

Reliable theoretical predictions of target-specific transition rates are important not only for current experiments, but also for planning the next generation of detectors. Compared to the DM-induced electron ionization rate in noble gases like xenon~\cite{Essig:2011nj,Graham:2012su,Lee:2015qva,Essig:2017kqs,Catena:2019gfa,Agnes:2018oej,Aprile:2019xxb,Aprile:2020tmw}, calculations for the DM-electron scattering rate in a crystal are more complicated. Ionization rates for noble gases can be calculated by considering each noble gas atom as an individual target, where the calculation simplifies to finding the ionization rate from an isolated atom, for which the wave functions and energy levels are well tabulated~\cite{Bunge:1993jsz}. However, for crystal targets the atoms are not isolated and more involved techniques are required to obtain an accurate characterization of DM-electron interactions in a many-body system.

There have been a variety of approaches taken to compute the DM-electron scattering rate in crystals. One of the first attempts, Ref.~\cite{Graham:2012su}, computed the rate with semi-analytic approximations for the initial and final state wave functions, and used the density of states to incorporate the electronic band structure. Later, Ref.~\cite{Lee:2015qva} continued in this direction and used improved semi-analytic approximations for the initial state wave functions. Meanwhile, a fully numerical approach was advanced in Refs.~\cite{Essig:2011nj, Essig:2015cda, Derenzo:2016fse} where density functional theory (DFT) was employed to calculate the valence and conduction electronic band structures and wave functions. The latter approach, as implemented in the \texttt{QEdark} program and embedded in the \textsc{Quantum ESPRESSO} package~\cite{QE-2009,QE-2017,doi:10.1063/5.0005082}, has become the standard for first-principles calculations of DM detection rates. Recently, in Refs.~\cite{Trickle:2019nya, Griffin:2019mvc} we used a similar DFT approach as implemented in our own program for a study of DM-electron scattering in a variety of target materials. More recently there has been work utilizing the relation between the dielectric function and the spin-independent scattering rate~\cite{Hochberg:2021pkt,Knapen:2021run,Knapen:2021bwg}, which also properly incorporates screening effects. 

The goal of this work is to further extend the DM-electron scattering calculation in several key aspects, and present state-of-the-art predictions for Si and Ge detectors using a combination of DFT and semi-analytic methods. 
As we will elaborate on shortly, the time- and resource-consuming nature of DFT calculations presents an intrinsic difficulty that has limited the scope of previous work in this direction to a restricted region of phase space; typically only bands within a few tens of eV above and below the band gap were included and electronic wave functions were cut off at a finite momentum.
We overcome this difficulty by implementing all-electron (AE) reconstruction (whose importance was previously emphasized in Ref.~\cite{Liang:2018bdb}) to recover higher momentum components of DFT-computed wave functions, and by extending the calculation to bands farther away from the band gap using semi-analytic approximations along the lines of Refs.~\cite{Graham:2012su,Lee:2015qva}. 
As we will see, the new contributions computed here have a significant impact on detection prospects in cases where higher energy and/or momentum regions of phase space dominate the rate, including scattering via a heavy mediator, and experiments with $\mathcal{O}(10\,\text{eV})$ or higher energy thresholds. 
We also stress that in contrast to the recent work emphasizing the relation between spin-independent DM-electron scattering rates and the dielectric function~\cite{Hochberg:2021pkt,Knapen:2021run,Knapen:2021bwg}, 
our calculation can be straightforwardly extended to DM models beyond the standard spin-independent coupling. 
Furthermore, we do not make assumptions about isotropy for the majority of our calculation, and our framework is capable of treating anisotropic targets which exhibit smoking-gun daily modulation signatures~\cite{Coskuner:2019odd,Trickle:2019nya,Geilhufe:2019ndy} (see also Refs.~\cite{Griffin:2018bjn,Coskuner:2021qxo} for discussions of daily modulation for phonon excitations).

Our calculation is implemented in an open-source program \texttt{EXCEED-DM} (EXtended Calculation of Electronic Excitations for Direct detection of Dark Matter), to be released in an upcoming publication. 
Currently a beta version of the program is available \href{https://github.com/tanner-trickle/EXCEED-DM}{here}~\cite{tanner_trickle_2021_4747696}. 
We also make available our DFT-computed wave functions~\cite{Trickle2021} and the output of \texttt{EXCEED-DM}~\cite{Trickle2021a} for Si and Ge.

\subsection{Overview of the Calculation and Key Results}
\label{subsec:key_results}

\begin{figure}[t]
	\centering
	\includegraphics[width=0.4\textwidth]{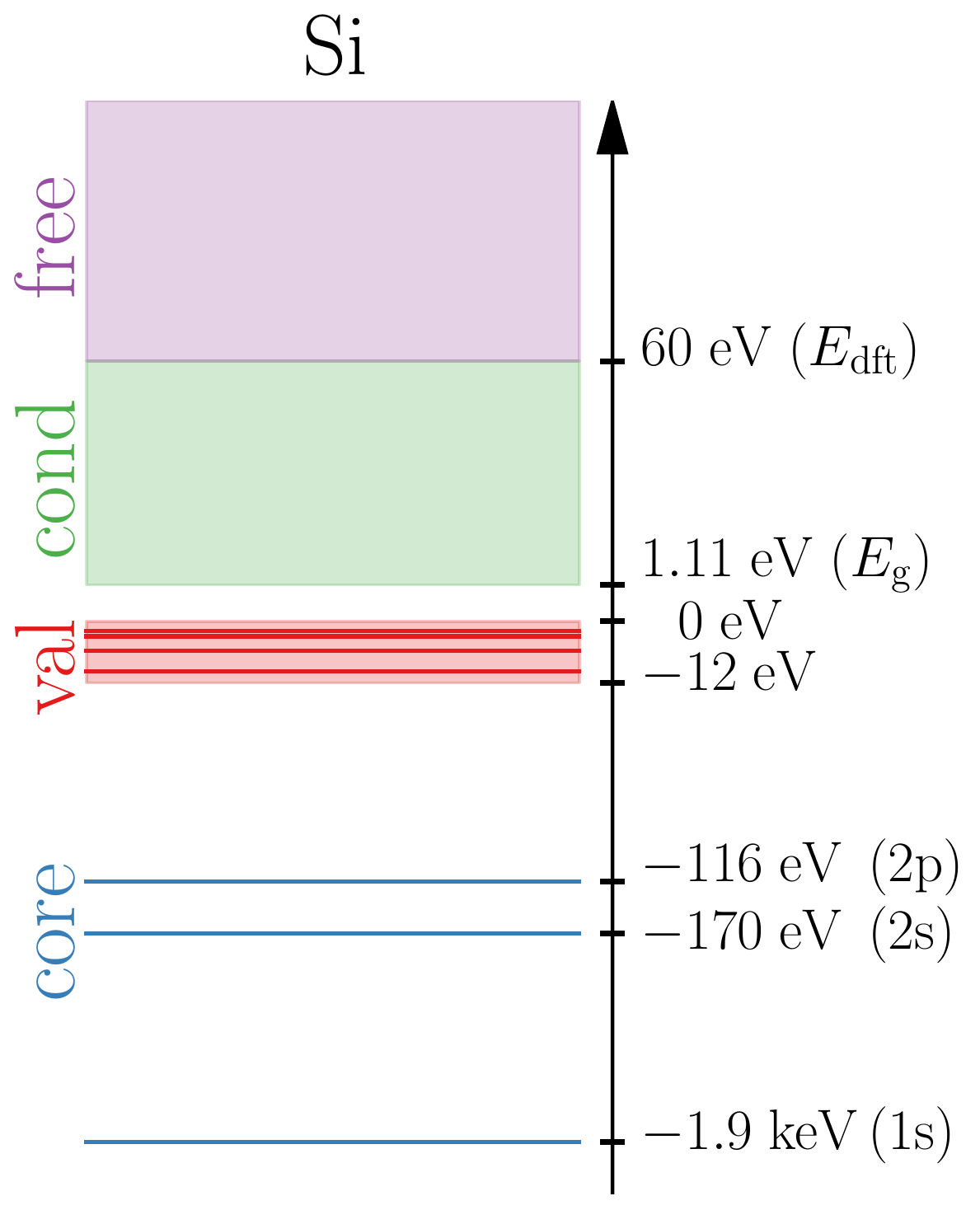}
	\hspace{0.05\textwidth}
	\includegraphics[width=0.4\textwidth]{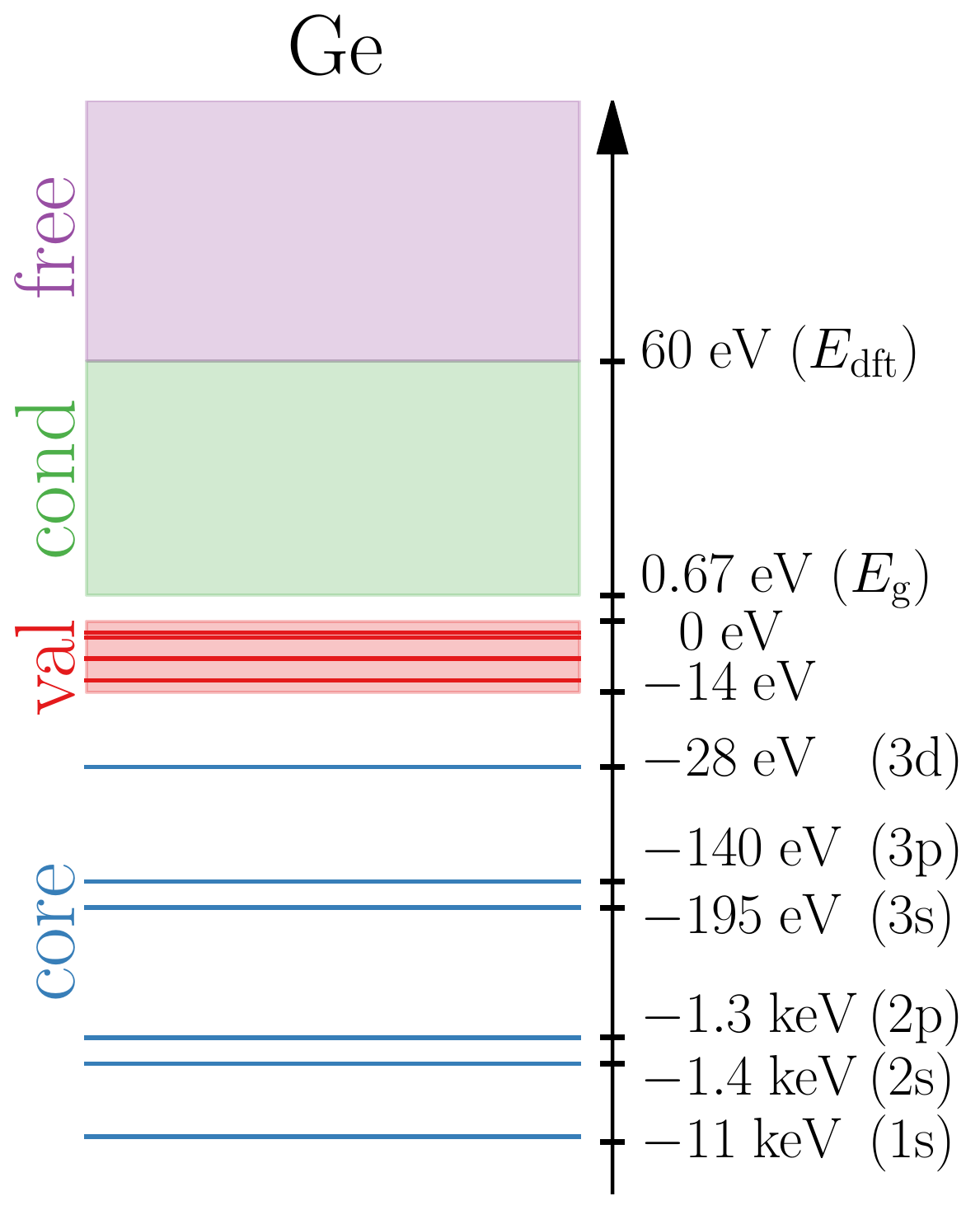}
    \caption{Schematic representation of electronic states in Si (left) and Ge (right), divided into core, valence (``val''), conduction (``cond'') and free. Shaded regions indicate the range of energies for each type of electronic states. In a scattering process, electrons transition from either core or valence states, below the Fermi surface at $E = 0$, to conduction or free states above the band gap $E_\text{g}$. As outlined in Sec.~\ref{subsec:key_results} and explained in detail in Sec.~\ref{sec:elec_states}, we compute the valence and conduction states numerically using DFT (including all-electron reconstruction), model the core states semi-analytically with RHF wave functions, and treat the free states as plane waves.}
	\label{fig:band_cartoon}
\end{figure}

\begin{figure}[t]
	\centering
	\includegraphics[width=\textwidth]{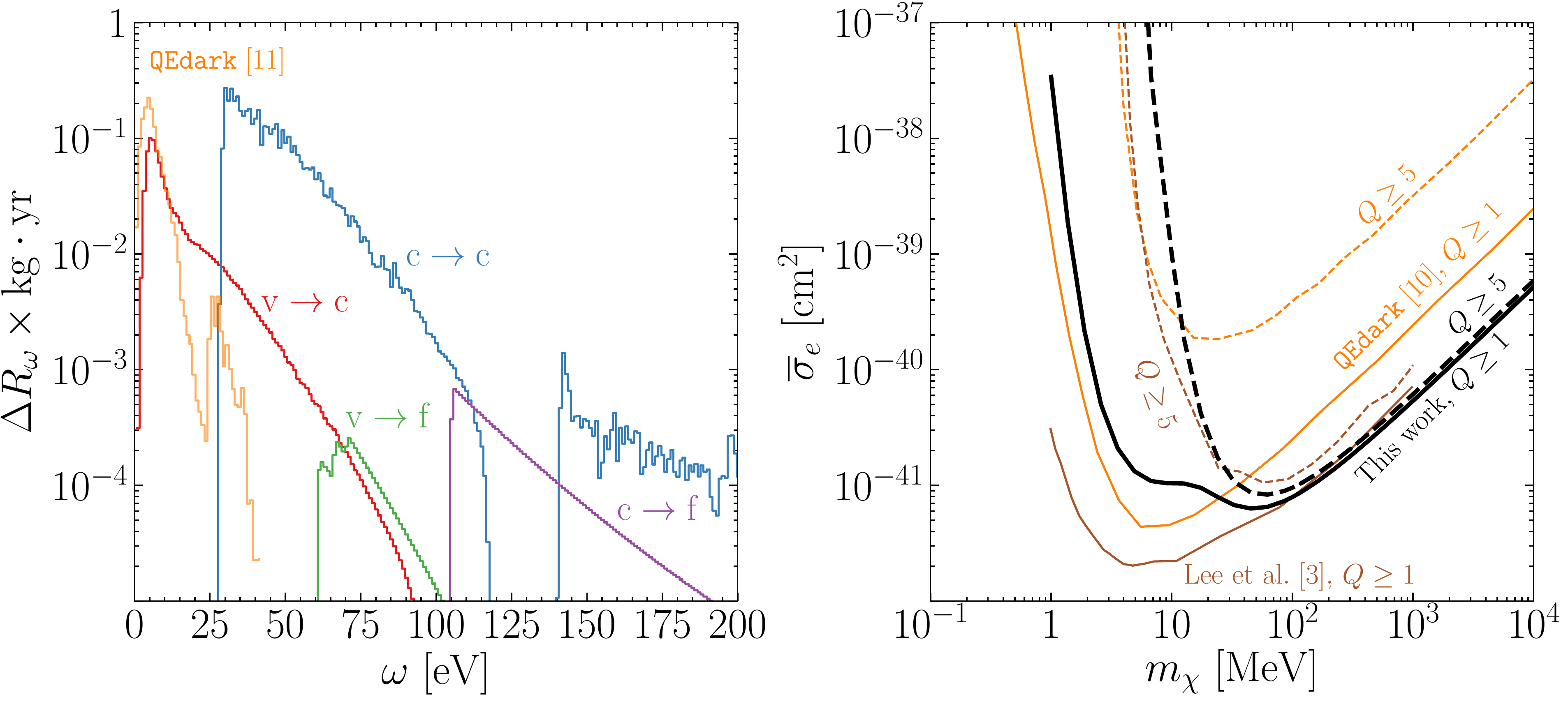}
	\caption{\label{fig:summary_plot}
		Selection of results from Sec.~\ref{sec:projected_constraints}, for DM-electron scattering via a heavy mediator in a Ge target. 
    \textbf{Left:} Contribution from each of the four transition types, valence to conduction (v$\to$c), valence to free (v$\to$f), core to conduction (c$\to$c), and core to free (c$\to$f) to the scattering rate binned in energy deposition (with $\Delta\omega=1$\,eV) for a 1\,GeV DM at a given reference cross section $\overline{\sigma}_e = 10^{-40}\; \text{cm}^2$. \textbf{Right:} 95$\%$ C.L.\ projected limit (3 events) on $\overline{\sigma}_e$ assuming 1\,kg-year exposure, for energy thresholds corresponding to 1 and 5 electron-hole pairs. We compare our results with \texttt{QEdark} calculations in Refs.~\cite{Essig:2015cda,Derenzo:2016fse} and the semi-analytic model of Lee et al~\cite{Lee:2015qva}; see text for details.}
\end{figure}

Before delving into the technical details, let us give a brief overview of the calculation and highlight some key results. 
We divide the electronic states in a (pure) crystal into four categories: core, valence, conduction and free, as illustrated in Fig.~\ref{fig:band_cartoon} for Si and Ge and discussed in more detail in Sec.~\ref{sec:elec_states}. 
At zero temperature, electrons occupy states up to the Fermi energy, defined as the maximum of the valence bands and denoted by $E=0$. 
The band gap $E_\text{g}$, {\it i.e.}\ the energy gap between the occupied valence bands and unoccupied conduction bands, is typically $\mathcal{O}(\text{eV})$ for semiconductors, {\it e.g.}\ 1.11\,eV for Si and 0.67\,eV for Ge; this sets a lower limit on the energy deposition needed for an electron transition to happen.

The electronic states near the band gap deviate significantly from atomic orbitals and need to be computed numerically. We apply DFT methods (including AE reconstruction) for this calculation, and refer to the DFT-computed states as valence and conduction. 
Specifically, for both Si and Ge, we take the first four bands below the gap to be valence, which span an energy range of $-12\,$eV to 0 and $-14\,$eV to 0, respectively, and take bands above the gap up to $E_\text{dft}=60\,$eV to be conduction. 

With more computing power we can in principle include more states, both below and above the band gap, in the DFT calculation. 
However, since the states far from the band gap can be modeled semi-analytically with reasonable accuracy, computing them with DFT is inefficient. 
Below the valence bands, electrons are tightly bound to the atomic nuclei. We model them using semi-analytic atomic wave functions and refer to them as core states. 
These include the 1s, 2s, 2p states in Si and 1s, 2s, 2p, 3s, 3p, 3d states in Ge (the 3d states in Ge are sometimes referred to as semi-core, and we will compare the DFT and semi-analytic treatment of them in Secs.~\ref{subsec:atomic_wf} and \ref{subsec:cc}). 
Finally, above $E_\text{dft}=60\,$eV, we model the states as free electrons as they are less perturbed by the crystal environment.

With the electronic states modeled this way, we compute the rate for valence to conduction (v$\to$c), valence to free (v$\to$f), core to conduction (c$\to$c) and core to free (c$\to$f) transitions induced by DM scattering, as discussed in detail in Sec.~\ref{sec:general_calc}. 
The total rate is the sum of all four contributions. 
We then use our calculation to update the projected reach of direct detection experiments in Sec.~\ref{sec:projected_constraints}, and compare our results with previous literature.

Figure~\ref{fig:summary_plot} gives a glimpse of some of our key results. 
Here we consider the case of DM scattering via a heavy mediator in a Ge target. 
The impact of core (3d) to conduction contributions is clearly visible from both the differential rate (left panel, for $m_\chi=1\,$GeV) and the projected reach (right panel). 
They dominate the total rate for $m_\chi\gtrsim 10\,$MeV, and, as we can see from the right panel of Fig.~\ref{fig:summary_plot}, lead to significantly more optimistic reach compared to previous DFT calculations implemented in \texttt{QEdark}~\cite{Essig:2015cda,Derenzo:2016fse}; this is especially true for higher detector thresholds (corresponding to higher $Q$ values). 
Note that while Refs.~\cite{Essig:2015cda,Derenzo:2016fse} included the 3d states in their DFT calculation, their contributions were significantly underestimated due to the absence of AE reconstruction. 
The importance of AE reconstruction is also seen from the valence to conduction differential rate in the left panel of Fig.~\ref{fig:summary_plot}, where our calculation predicts a much higher rate at $\omega \gtrsim 15\,$eV compared to the \texttt{QEdark} calculation in Ref.~\cite{Derenzo:2016fse}.
Meanwhile, accounting for in-medium screening (see Sec.~\ref{subsec:in-med}) we find, consistent with Ref.~\cite{Knapen:2021run}, a lower rate at energy depositions just above the band gap, and as a result, weaker reach at low $m_\chi$, compared to Refs.~\cite{Essig:2015cda,Derenzo:2016fse}.
On the other hand, our modeling of the core (3d) states is similar to the semi-analytic approach of Ref.~\cite{Lee:2015qva}, and indeed we find very similar reach at large $m_\chi$; however, the approach of Ref.~\cite{Lee:2015qva} overestimates the rate at smaller $m_\chi$ due to reduced accuracy in the modeling of the valence and conduction states. 
We reserve a more detailed comparison with the literature for Sec.~\ref{subsec:compare_to_prev}.

\section{Electronic States}
\label{sec:elec_states}

\begin{figure}[t]
    \centering
    \includegraphics[width=\textwidth]{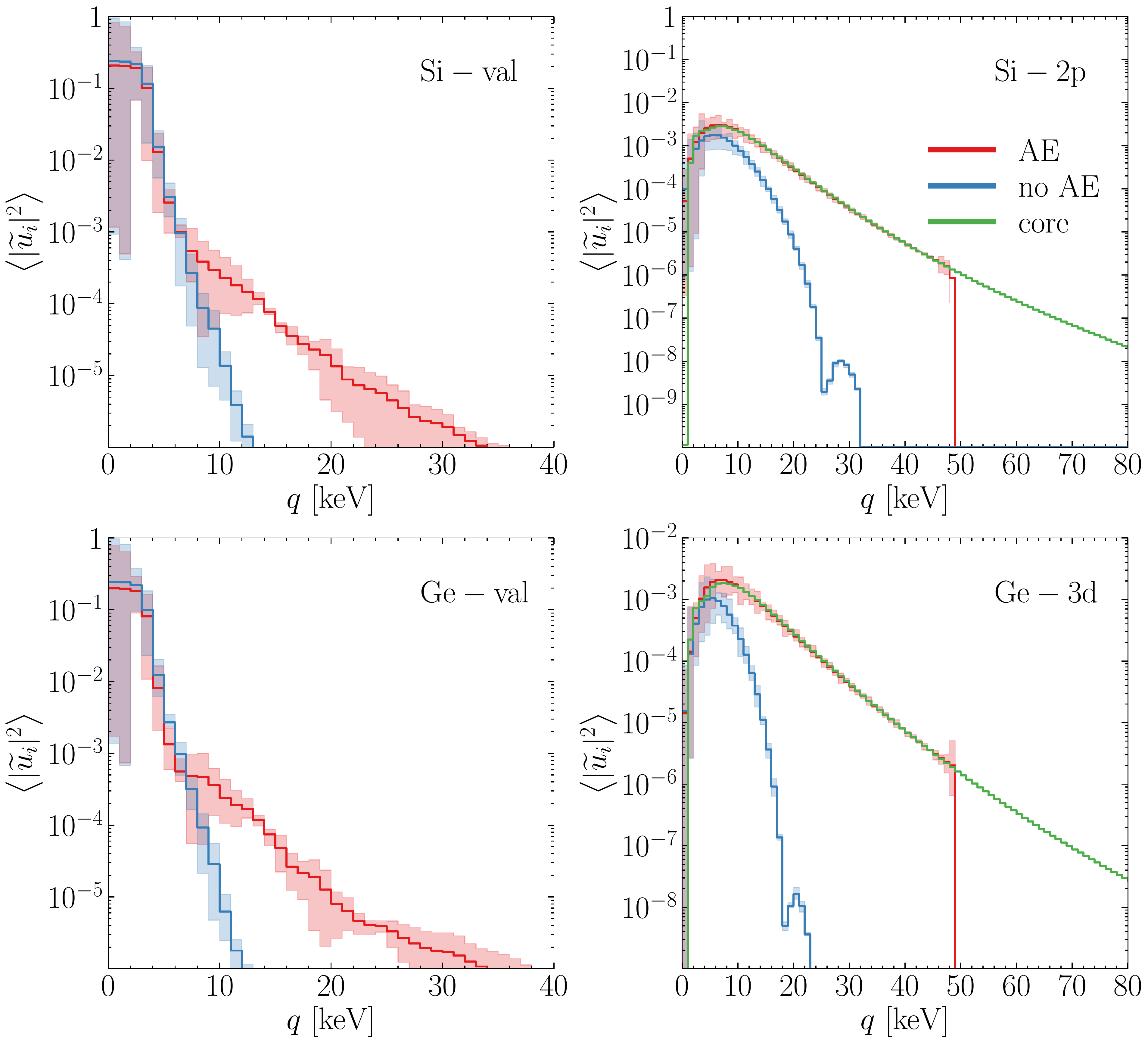}
    \caption{Comparison of the Bloch wave function magnitudes, defined in Eq.~\eqref{eq:av_bloch_state}, computed with DFT with (red, ``AE'') and without (blue, ``no AE'') AE reconstruction, and the semi-analytic core approximation of Eq.~\eqref{eq:u_jnlmk} (green, ``core''). Shaded bands indicate the maximum and minimum values across all the bands belonging to the state type indicated in the upper right corner of each panel. AE reconstruction, discussed in Sec.~\ref{subsubsec:ae}, recovers the large momentum behavior of the electronic wave functions. Core electronic states, such as those shown in the right panels and discussed in Sec.~\ref{subsec:atomic_wf}, can be well modeled semi-analytically with atomic wave functions, as seen by the good agreement between the ``core'' and ``AE'' curves. When applicable, the semi-analytic parameterization is advantageous since the electronic wave functions are then known to arbitrarily large momentum.}
    \label{fig:binned_core_wf}
\end{figure}

To compute the DM-electron scattering rate one must understand the electronic states of the target. In targets with a periodic potential, Bloch's theorem states that the energy eigenstates can be indexed by a momentum, $\vect{k}$, which lies within the first Brillouin zone (1BZ). These Bloch states, $\psi_{i,\vect{k}}$, where $i$ represents additional quantum numbers, are eigenstates of the discrete translation operator such that $\psi_{i,\vect{k}}(\vect{x} + \vect{r}) = e^{i \vect{k} \cdot \vect{r}} \psi_{i,\vect{k}}(\vect{x})$, which means the electronic wave functions can be written as
\begin{align}
    \label{eq:bloch_form}
    \psi_{i,\vect{k}}(\vect{x}) = \frac{1}{\sqrt{V}}\, e^{i \vect{k} \cdot \vect{x}} \,u_{i,\vect{k}}(\vect{x}) \,,
\end{align}
where $u_{i,\vect{k}}(\vect{x} + \vect{r}) = u_{i,\vect{k}}(\vect{x})$ and $V$ is the target volume. For every $\vect{k}$ there exists a tower of eigenstates (labeled by $i$) of the target Hamiltonian which constitutes the complete set of states in the target. Unfortunately this complete set is not known for a general material and therefore a combination of approximations must be used to calculate them. 
As discussed in Sec.~\ref{subsec:key_results} and illustrated in Fig.~\ref{fig:band_cartoon}, we divide the states into core, valence, conduction and free, and use a combination of numerical calculations and semi-analytic modeling. 
In this section, we expand on the treatment of each type of electronic states.  

We first discuss the DFT calculation for valence and conduction states in Sec.~\ref{subsec:dft}, and then move on to explain the semi-analytic treatment of core states in Sec.~\ref{subsec:atomic_wf}. Our main results are contained in Fig.~\ref{fig:binned_core_wf} where we compare the average magnitude of electronic wave functions,
binned in momentum (see Eq.~\eqref{eq:av_bloch_state}), computed with and without AE reconstruction, discussed further in Sec.~\ref{subsubsec:ae}, and, for the highest energy core states (2p in Si and 3d in Ge), those computed using the core approximation discussed in Sec.~\ref{subsec:atomic_wf}. 
We find that the AE reconstruction includes a significant contribution from wave functions at large momentum as expected, and that for the core states, the semi-analytic approach reproduces the large momentum components of these AE reconstructed DFT wave functions. 
Lastly we will discuss the analytic treatment of the free states in Sec.~\ref{subsec:pw_approx}.

\subsection{DFT Wave Functions and Band Structures}
\label{subsec:dft}

\begin{figure}[t]
\centering
\includegraphics[width=0.48\textwidth]{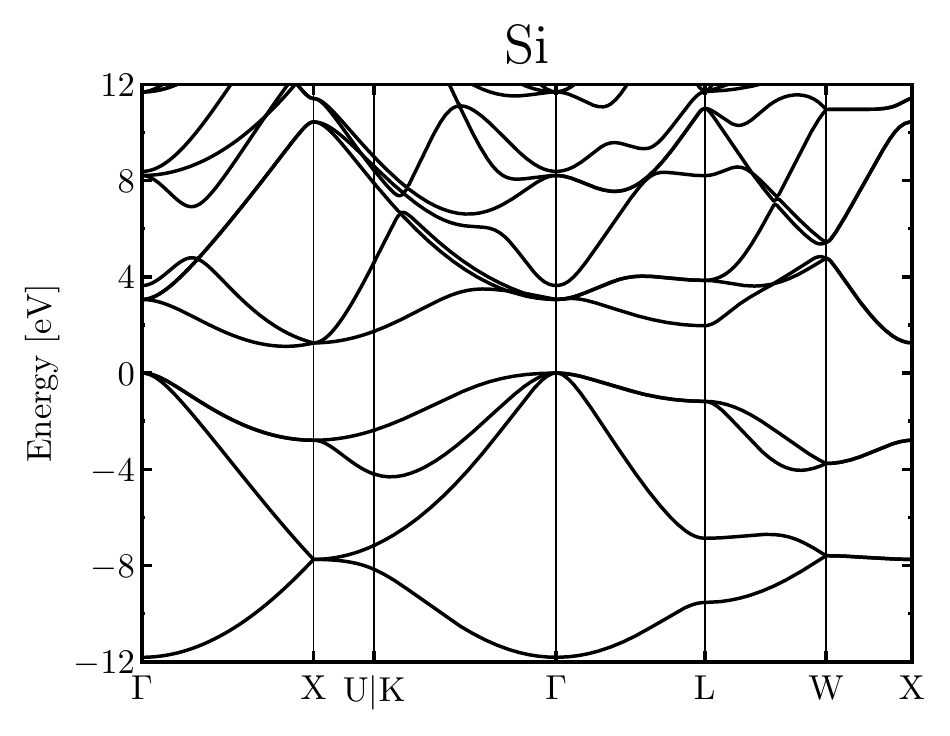}
\includegraphics[width=0.48\textwidth]{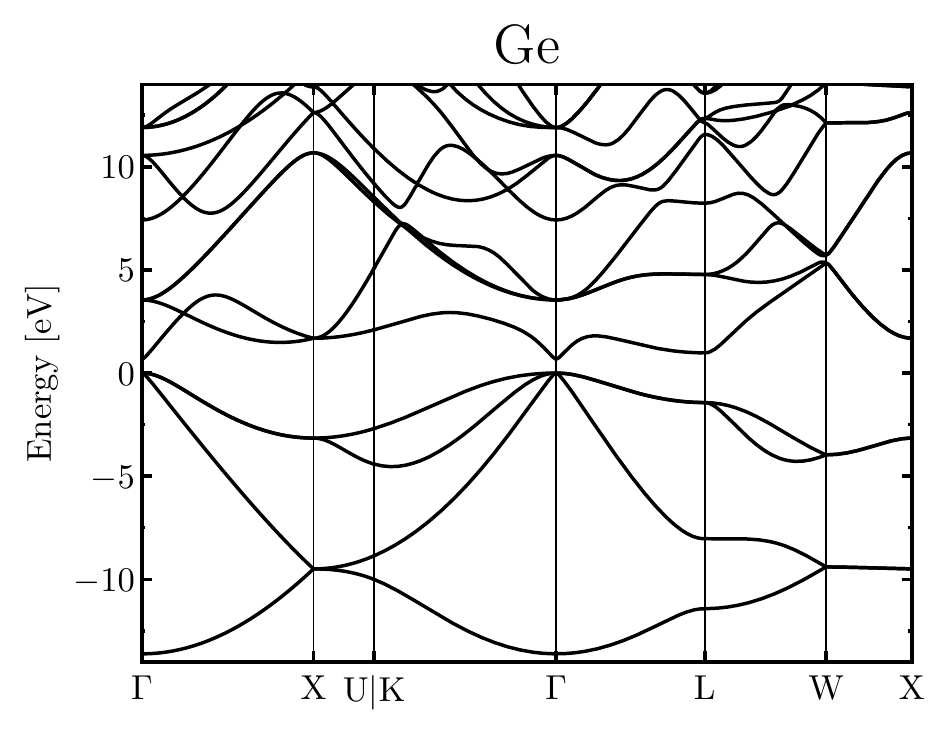}
\caption{Calculated band structures of Si (left) using a PBE xc-functional within DFT and Ge (right) using a hybrid functional HSE06. The band gaps have been scissor corrected to their measured values near zero temperature, 1.11\,eV and 0.67\,eV for Si and Ge, respectively. The Fermi level is set to 0\,eV in both panels.}
\label{fig:bandstructures}
\end{figure}

In principle, DFT provides an exact solution to the many-electron Schr\"odinger equation by the Hohenberg-Kohn theorems that treat all properties of a quantum many-body system as unique functionals of the ground state density. 
They further show that the exact ground state density and energy can be found by minimizing the total energy of the system~\cite{Hohenberg:1964zz, martin2020electronic}. This becomes tractable by the Kohn-Sham (KS) equations that reduce the many-body problem to non-interacting electrons moving in an effective potential, $V_{\mathrm{eff}}$, 
\begin{equation}
\left( \frac{p^2}{2m_e} + V_{\mathrm{eff}} - \epsilon_i \right) \psi_i = 0 \,,\qquad\quad
V_{\mathrm{eff}} = V_{\mathrm{ext}} + V_{\mathrm{H}} + V_{\mathrm{xc}}\,,
\label{eq:dft_hamiltonian_eq}
\end{equation}
where $\epsilon_i$ is the orbital energy of the KS orbital $\psi_i$~\cite{Kohn:1965zzb}. The external potential $V_{\mathrm{ext}}$ and Hartree potential $V_{\mathrm{H}}$, are known, while the exchange-correlation (xc) potential $V_{\mathrm{xc}}$, which contains the many-body interactions, must be approximated. Herein lies the deviation from the exact solution, and although various formulations of xc-energy functionals have been successful, the choice of xc-functional will affect the predicted electronic states and hence calculated transition rates. For Si, we use PBE~\cite{Perdew1996}, a type of generalized gradient approximation (GGA) xc-functional which is one of the most popular and low-cost choices. Local and semi-local based xc-functionals, such as PBE, suffer from a self-interaction error and band gap underestimation, which we modify with a ``scissor correction'' where the bands are shifted to match the experimentally determined values of band gap. For Ge, this underestimation results in zero band gap with PBE, therefore we instead use a hybrid functional, which mixes a parameterized amount of exact exchange into the xc-functional, correcting band gaps and band widths by error cancellation at the cost of increased computation time. We use the range-separated hybrid functional HSE06~\cite{Heyd2003a,Heyd2006}, which applies a screened Coulomb potential to correct the long-range behavior of the xc-potential, giving high accuracy at a mid-level computational cost. Our computed band structures for Si and Ge are shown in Fig.~\ref{fig:bandstructures}.

The periodic Bloch wave functions, $u_{i,\vect{k}}(\vect{x})$, Eq.~\eqref{eq:bloch_form}, for band $i$ and Bloch wave vector $\vect{k}$ are computed by finding the Fourier coefficients, $\widetilde{u}_{i, \vect{k}, \vect{G}}$ (which satisfy the normalization condition, $\sum_\vect{G} | \widetilde{u}_{i, \vect{k}, \vect{G}}|^2 = 1$):
\begin{align}
    u_{i, \vect{k}}(\vect{x}) = \sum_\vect{G} \widetilde{u}_{i,\vect{k},\vect{G}} \,e^{i \vect{G} \cdot \vect{x}} \, .
    \label{eq:bloch_fourier}
\end{align}
The number of reciprocal lattice vectors $\vect{G}$ kept in the sum is conventionally set by an energy cutoff, $E_\text{cut}$, such that $|\vect{k} + \vect{G}|^2 < 2 m_e E_\text{cut}$. These Bloch wave function coefficients $\widetilde{u}_{i,\vect{k},\vect{G}}$ for both Si and Ge are computed with the projector augmented wave (PAW) method~\cite{Blo,Kresse1999} within \textsc{vasp}~\cite{Kresse1993,Kresse1994,Kresse1996,Kresse1996a} up to $E_\text{cut} = 1$ keV on a $10\times10\times10$ uniform $\vect{k}$ mesh over the 1BZ. We then include AE reconstruction effects up to a higher energy cutoff, $E_\text{AE} = 2$ keV, which recovers higher momentum components of the wave functions up to $|\vect{k} + \vect{G}|^2 < 2 m_e E_\text{AE}$, as discussed in more detail in Sec.~\ref{subsubsec:ae}. The Bloch wave function coefficients, $\widetilde{u}_{i,\vect{k},\vect{G}}$, for Si and Ge used for this work can be found here~\cite{Trickle2021}.

A final consideration of using DFT wave functions is that DFT is fundamentally a ground state method, and the KS conduction band states are only approximations to excited states. Excited state methodologies are much more computationally expensive than ground state KS-DFT. Furthermore, since excited state quasiparticles, such as excitons, have been argued to have a negligible effect on the calculation of DM scattering rates~\cite{Derenzo:2016fse}, they are neglected in our calculations. 

\subsubsection{All-electron Reconstruction}
\label{subsubsec:ae}

There are many different approaches to find the eigenstates of Eq.~\eqref{eq:dft_hamiltonian_eq}. The PAW method~\cite{Blo} is one such standard approach. The main idea of the PAW method is to split up the calculation of the eigenstates: near the ionic centers the wave functions resemble the eigenstates of an isolated atom, while further away they can be computed numerically with a pseudopotential. This greatly simplifies the numeric calculation since the focus is then on large distance (small momentum), and the small distance (high momentum) pieces can be self-consistently reintroduced after the main part of the DFT calculation. The large distance components of the wave function are known as ``pseudo wave functions'' (PS wave functions) and the total wave functions are known as the ``all-electron wave functions'' (AE wave functions), indicating that all of the wave function components are included. We will now give a brief overview of how the AE wave functions can be reconstructed from the PS wave functions, computed with PAW-based DFT codes, and refer the reader to Refs.~\cite{Blo,Kresse1999,rostgaard2009projector,pawpy} for more detailed information.\footnote{It is possible to calculate all electronic eigenstates, including the core, self-consistently by other more complex methods such as the full-potential linearized augmented plane wave (FP-LAPW) method or the relaxed-core PAW (RC-PAW) method.} 

The AE wave functions, $| \Psi^\text{AE} \rangle$ are built from two components. Near the ionic core, or inside an ``augmentation sphere,'' $| \Psi^\text{AE} \rangle$ is expanded in a set of basis functions, $| \phi^\text{AE}_j \rangle$, which are simply taken to be the wave functions of an isolated atom,
\begin{align}
    | \Psi^\text{AE} \rangle = \sum_j c_j \,| \phi^\text{AE}_j \rangle \, .
\end{align}
Outside of the augmentation sphere, $| \Psi^\text{AE} \rangle = | \Psi^\text{PS} \rangle$. Near the ionic core the PS wave functions $| \Psi^\text{PS} \rangle$ are expanded in a set of basis functions $| \phi^\text{PS}_j \rangle$ that are computationally more convenient than the $| \phi^\text{AE}_j \rangle$. Therefore,
\begin{align}
    | \Psi^\text{AE} \rangle = | \Psi^\text{PS} \rangle - \sum_j c_j' \,| \phi^\text{PS}_j \rangle + \sum_j c_j \,|\phi^\text{AE}_j \rangle \, ,
\end{align}
which simply replaces the components in $| \Psi^\text{PS} \rangle$ within the augmentation sphere with the AE wave function. To find the $c$ coefficients we insert an identity, $\identity = \sum_j | \phi^\text{AE}_j \rangle \langle p^\text{AE}_j | = \sum_j | \phi^\text{PS}_j \rangle \langle p^\text{PS}_j |$, where $|p^{\text{AE}/\text{PS}}_j \rangle$ are projector functions, defined to satisfy this identity within the augmentation sphere. Therefore, $c_j = \langle p^\text{AE}_j | \Psi^\text{AE} \rangle$, $c_j' = \langle p^\text{PS}_j | \Psi^\text{PS} \rangle$. The last ingredient to compute $| \Psi^\text{AE} \rangle$ from $| \Psi^\text{PS} \rangle$ is to require that $| \phi^\text{AE}_j \rangle$ is related to $| \phi^\text{PS}_j \rangle$ via a transformation, $| \phi^\text{AE} \rangle = \mathcal{T} | \phi^\text{PS} \rangle$. This implies that all the PS states are related to AE states by this transformation $\mathcal{T}$, such that $c_j = c_j'$ and the AE reconstruction can be written as
\begin{align}
    | \Psi^\text{AE} \rangle = | \Psi^\text{PS} \rangle + \sum_j \left( | \phi^\text{AE}_j \rangle - | \phi^\text{PS}_j \rangle \right) \langle p^\text{PS}_j | \Psi^\text{PS} \rangle \, .
\end{align}
In practice, we implement the AE reconstruction with \texttt{pawpyseed}~\cite{pawpy}, and the plane wave expansion cutoff of $| \Psi^\text{AE} \rangle$, $E_\text{AE}$, can be increased from the initial $E_\text{cut}$. We use $E_\text{AE} = 2$ keV.

To visualize the effect of AE reconstruction, we plot in Fig.~\ref{fig:binned_core_wf} the average magnitude of the Bloch wave functions, binned in $q$,
\begin{align}
    \left\langle \left| \widetilde{u}_i \right|^2 \right\rangle (q; \Delta q) \equiv \frac{1}{N_q} \sum_{\vect{k}}\sum_{\vect{G}} |\widetilde{u}_{i,\vect{k},\vect{G}}|^2 \,\theta(q + \Delta q - |\vect{k} + \vect{G}|) \,\theta(|\vect{k} + \vect{G}| - q) \,,
    \label{eq:av_bloch_state}
\end{align}
where $\widetilde{u}_{i,\vect{k},\vect{G}}$ are the Fourier components of the Bloch wave functions, defined in Eq.~\eqref{eq:bloch_fourier}. Each bin in momentum space extends from $q$ to $q + \Delta q$ with $\Delta q=1\,$keV, and $N_q$ is a normalization factor equal to the number of points in a bin, $N_q = \sum_{\vect{k}} \sum_{\vect{G}} \theta(q + \Delta q - |\vect{k} + \vect{G}|)\,\theta(|\vect{k} + \vect{G}| - q)$. 
We see that AE reconstruction recovers the high momentum components, which as we will see can significantly affect the DM-induced transition rate for processes which favor large momentum transfers (such as processes mediated by heavy particles), or processes limited to larger $\omega$ ({\it e.g.}\ higher experimental thresholds where large $q$ processes are the only kinematically allowed transitions).
Previous DFT calculations of DM-induced electron transition rates, with the exceptions of Ref.~\cite{Liang:2018bdb, Trickle:2019nya, Griffin:2019mvc}, used only the pseudo wave functions, $| \Psi^\text{PS} \rangle$ as opposed to the AE wave functions, $| \Psi^\text{AE} \rangle $, and have therefore underestimated detection rates in several cases.

\subsection{Atomic Wave Functions}
\label{subsec:atomic_wf}

If one could reconstruct the AE wave functions arbitrarily deep into the band structure, and to arbitrarily high momentum, one could calculate an accurate representation of the complete set of electronic states with a DFT calculation. In practice, however, this is neither feasible nor necessary. States deep in the band structure are more isolated from the influence of the crystal environment, and so an isolated atomic approximation becomes valid. We refer to these inner, tightly bound electrons as core electrons. In Si, we will show that the 2p states and below can be treated as core, while in Ge, the 3d states and below can, as alluded to in Fig.~\ref{fig:band_cartoon}. The purpose of this subsection is to expand on the atomic approximation for core electrons and discuss its accuracy. 

More precisely, the initial states of a transition should be taken as a linear combination of isolated atomic wave functions that is in Bloch form (known as Wannier states):
\begin{align}
    \label{eq:core_elec_wf}
    \psi_{\kappa nlm,\vect{k}}(\vect{x}) = \frac{1}{\sqrt{N}} \sum_\vect{r} e^{i \vect{k} \cdot (\vect{r}+\vect{x}_\kappa)} \,\psi_{\kappa nlm}^\text{atom}(\vect{x} - \vect{r} - \vect{x}_\kappa) \,,
\end{align}
where $\kappa$ labels the atom in the primitive cell, $n, l, m$ are the standard atomic quantum numbers, $\vect{x}_\kappa$ is the equilibrium position of the $\kappa^\text{th}$ atom, $\sum_\vect{r}$ sums over all primitive cells in the lattice, and $N$ is the total number of cells. 
In contrast to the valence and conduction states discussed in the previous subsection, the core states are labeled by $(\kappa nlm)$ rather than band index $i$. 
The corresponding periodic (dimensionless) $u$ functions can be easily obtained via Eq.~\eqref{eq:bloch_form}:
\begin{equation}
u_{\kappa nlm,\vect{k}}(\vect{x}) = \sqrt{\Omega}\, \sum_\vect{r} e^{-i \vect{k} \cdot (\vect{x}-\vect{r}-\vect{x}_\kappa)} \,\psi_{\kappa nlm}^\text{atom}(\vect{x} - \vect{r} - \vect{x}_\kappa) \,,
\label{eq:u_jnlmk}
\end{equation}
where $\Omega=V/N$ is the primitive cell volume. 

In general, the atomic wave functions $\psi_{\kappa nlm}^\text{atom}$ are not known analytically, but are expanded in a basis of well-motivated analytic functions. The basis coefficients are then fit by solving the isolated atomic Hamiltonian, giving a semi-analytic expression for $\psi_{\kappa nlm}^\text{atom}$. We use a basis of Slater type orbital (STO) wave functions whose radial component is
\begin{align}
    R_\text{STO}(r; Z, n) & = a_0^{-3/2} \frac{\left(2 Z \right)^{n + \frac{1}{2}}}{\sqrt{(2n)!}}  \left( \frac{r}{a_0} \right)^{n - 1} e^{-Z r/a_0} \,, 
   \label{eq:sto_wf}
\end{align}
where $a_0 =0.53$\,\AA\;$= (3.7\,\text{keV})^{-1}$ is the Bohr radius, and $Z$ is an effective charge of the ionic potential. Including the angular part, the atomic wave functions are then
\begin{align}
    \psi^{\rm atom}_{\kappa nlm}(\vect{x}) = \sum_j C_{jln,\kappa} R_\text{STO}(x; Z_{jl,\kappa}, n_{jl,\kappa}) Y_l^m(\hat{\vect{x}}) \,,
\end{align}
where $C_{jln,\kappa}, Z_{jl,\kappa}, n_{jl,\kappa}$ are tabulated in Ref.~\cite{Bunge:1993jsz}, and $Y_l^m(\hat{\vect{x}})$ are the spherical harmonics with the Condon-Shortley phase convention~\cite{E.U.Condon1935}.

To assess the accuracy of the atomic wave function approximation, we temporarily push the DFT calculation beyond its default regime (valence and conduction), to the highest core states -- 2p states in Si and 3d states in Ge, where it is still computationally feasible -- and compare the numerical wave functions to the semi-analytic ones discussed above. 
The results, in terms of the average magnitude of Bloch wave functions defined in Eq.~\eqref{eq:av_bloch_state}, are shown in the right panels of Fig.~\ref{fig:binned_core_wf}.\footnote{The flatness of band structures offers a complementary check of the validity of the atomic approximation. We have verified that the DFT computed energy eigenvalues indeed have a small variance for the highest core states, as expected.}
We see that the atomic approximation accurately reproduces the numerical wave functions up to the momentum cutoff $\sqrt{2m_e E_\text{AE}}\simeq 50\,$keV for $E_\text{AE}=2\,$keV. 
These plots also show the limitation of DFT calculations. 
While AE reconstruction recovers higher-momentum components of electronic wave functions, it is not feasible to expand the plane wave basis set to arbitrarily high cutoff. 
However, having verified the atomic approximation for the highest core states, we can use it for all core states with confidence, allowing us to more easily include the high momentum components beyond the DFT cutoff.

\subsection{Plane Wave Approximation}
\label{subsec:pw_approx}

With the inclusion of the semi-analytic core states, all of the states below the band gap have been modeled. States above the band gap can also be computed with DFT methods, as described in Sec.~\ref{subsec:dft}. Similar to valence bands, there are practical limitations to how many conduction bands can be included. To remedy this in the simplest way possible, we model states far above the band gap as plane waves,
\begin{equation}
    \label{eq:free_elec_wf}
    \psi_{\vect{G}, \vect{k}}(\vect{x})  = \frac{1}{\sqrt{V}}\, e^{i \left( \vect{k} + \vect{G} \right) \cdot \vect{x}} \,,\qquad\quad
 E_{\vect{G}, \vect{k}}   = \frac{| \vect{k} + \vect{G} |^2}{2 m_e}\,,
\end{equation}
where $\vect{G}$ is a reciprocal lattice vector, and plays the role of a band index. (To understand this, simply note that every momentum can be decomposed into a $\vect{k}$ vector inside the 1BZ and a reciprocal lattice vector. Integrating over the momentum of plane wave states amounts to a $\vect{k}$ integral within the 1BZ and a $\vect{G}$ sum.)
From Eq.~\eqref{eq:bloch_form} we see that the corresponding periodic $u$ functions are simply
\begin{equation}
u_{\vect{G},\vect{k}}(\vect{x}) = e^{i\vect{G}\cdot\vect{x}} \,.
\label{eq:u_free}
\end{equation}
The plane wave approximation is often used in atomic ionization rate calculations, with the inclusion of a \textit{Fermi factor}, $F(\nu)$,
\begin{align}
    \label{eq:fermi_factor}
    F(\nu)  = \frac{\nu}{1 - e^{-\nu}}\,, \qquad \nu(Z_\text{eff}, E)  = 2 \pi Z_\text{eff} \frac{\alpha m_e}{\sqrt{2 m_e E}} \, , 
\end{align}
where $E$ is the final state electron energy, and $Z_\text{eff}$ is an effective charge parameter, which enhances the transition rate at low $E$ to account for the long range behavior of the Coulomb potential. See Refs.~\cite{Essig:2011nj,Graham:2012su, Lee:2015qva, Agnes:2018oej} for more details. In atomic ionization calculations one usually takes $Z_\text{eff}$ to be related to the binding energy of the initial state, $E_B$,
\begin{align}
    \label{eq:z_eff_binding_approx}
    Z_\text{eff} = n \sqrt{\frac{E_B}{13.6 \text{ eV}}} \, ,
\end{align}
where $n$ is the principal quantum number. Since the rate is proportional to the Fermi factor, $Z_\text{eff} = 1$ is seen as the conservative choice. Later in Secs.~\ref{subsec:vf} and \ref{subsec:cf} we quantify how much of an effect this has on the transition rate. This uncertainty is only important for very high experimental thresholds, and generally we find that $Z_\text{eff} = 1$ leads to a smoother match (within an $\mathcal{O}(1)$ factor) to conduction band contributions from DFT calculations.

\section{Electronic Transition Rates}
\label{sec:general_calc}

We now present the DM-induced electron transition rate calculation. We begin with a general discussion and then in Secs.~\ref{subsec:vc}-\ref{subsec:cf} consider the four different transition types in turn: valence to conduction ($\mathrm{v} \rightarrow \mathrm{c}$), valence to free ($\mathrm{v} \rightarrow \mathrm{f}$), core to conduction ($\mathrm{c} \rightarrow \mathrm{c}$) and core to free ($\mathrm{c}\rightarrow\mathrm{f}$). Finally, in Sec.~\ref{subsec:in-med} we discuss the treatment of in-medium screening.

The general derivation has been discussed previously (see {\it e.g.}\ Refs.~\cite{Graham:2012su, Essig:2015cda,Catena:2019gfa, Trickle:2019nya, Liang:2018bdb}), and we repeat it here for completeness and clarity, as a variety of conventions have been used. Beginning with Fermi's Golden Rule, the transition rate between electronic states $\ket{i, s}$ and $\ket{f, s'}$ due to scattering with an incoming non-relativistic DM particle, $\chi$, with mass $m_\chi$, velocity $\vect{v}$, and spin $\sigma$ is given by
\begin{align}
    \label{eq:fermi}
    \Gamma_{i, s, \sigma \rightarrow f, s', \sigma'}(\vect{v}) = 2 \pi V \int \frac{d^3 q}{(2 \pi)^3}\, \big|\bra{\vect{p'}, \sigma'; f, s'} \,\delta\hat{H} \,\ket{\vect{p}, \sigma; i, s} \big|^2 \,\delta(E_{f, s'} - E_{i, s} - \omega_\vect{q})\,,
\end{align}
where $\ket{\vect{p}, \sigma; i, s} = \ket{\vect{p}, \sigma} \otimes \ket{i, s}$, $\vect{q}$ is the momentum deposited onto the target, $\vect{p} = m_\chi \vect{v}$, $\vect{p}' = \vect{p} - \vect{q}$, $\delta\hat{H}$ is the interaction Hamiltonian, $V$ is total volume of the target, and $\omega_\vect{q}$ is the energy deposition:
\begin{equation}
    \omega_\vect{q} = \frac{1}{2} m_\chi v^2 - \frac{\left( m_\chi \vect{v} - \vect{q} \right)^2}{2 m_\chi} = \vect{q} \cdot \vect{v} - \frac{q^2}{2 m_\chi} \, .
    \label{eq:w_kin}
\end{equation}
We assume that all quantum states are unit normalized. Modulo in-medium screening effects, discussed below in Sec.~\ref{subsec:in-med}, we can write Eq.~\eqref{eq:fermi} in terms of the standard QFT matrix element, defined with plane wave incoming and outgoing states, by inserting $\mathbbm{1} = V \sum_s \int \frac{d^3k}{(2\pi)^3} \ket{\vect{k},s} \bra{\vect{k},s}$ and using 
\begin{align}
    \label{eq:qft_m}
    \langle \vect{p}', \sigma'; \vect{k}', s' | \,\delta \hat H\, | \vect{p}, \sigma; \vect{k}, s \rangle \equiv \frac{(2\pi)^3}{V^2} \frac{\mathcal{M}_{\sigma' s'\sigma s}(\vect{p}', \vect{k}', \vect{p}, \vect{k})}{4 m_e m_\chi} \,\delta^{(3)}\left( \vect{p}' + \vect{k}' - \vect{p} - \vect{k}\right) \, .
\end{align}
We find\\[-4pt]
\begin{minipage}{\textwidth}
\begin{align}
    \label{eq:fermi_m}
    \Gamma_{i, s, \sigma \rightarrow f, s', \sigma'}(\vect{v}) = &\;\frac{2 \pi}{16 V m_e^2 m_\chi^2} \int \frac{d^3 q}{(2 \pi)^3} \,\delta(E_{f, s'} - E_{i, s} - \omega_\vect{q}) \nonumber\\
    &\times \left| \int \frac{d^3k}{(2\pi)^3} \,\mathcal{M}_{\sigma' s'\sigma s}(\vect{p} -\vect{q}, \vect{k}+\vect{q}, \vect{p}, \vect{k}) \,\widetilde{\psi}_f^*(\vect{k} + \vect{q}) \widetilde{\psi}_i(\vect{k}) \right|^2  \, ,
\end{align}
\end{minipage}
where $\widetilde{\psi}_i(\vect{k}) = \sqrt{V} \langle \vect{k} | i \rangle$. 

We will limit our analysis to matrix elements which only depend on $\vect{q}$, and assume that the electron energy levels are also spin independent, which allows the spin sums to be easily computed:
\begin{align}
    \label{eq:fermi_spin_av}
    \overline{\Gamma}_{i \rightarrow f} & \equiv \frac{1}{2} \sum_{\sigma, \sigma'} \sum_{s, s'} \Gamma_{i, s, \sigma \rightarrow f, s', \sigma'} \nonumber\\
    & = \frac{4 \pi}{16 V m_e^2 m_\chi^2}\int \frac{d^3 q}{(2 \pi)^3} \,\overline{|\mathcal{M}(\vect{q})|^2}\, |f_{i \rightarrow f}|^2 \,\delta \left( E_f - E_i - \omega_\vect{q} \right) , \\
    f_{i \rightarrow f} & \equiv \int \frac{d^3k}{(2\pi)^3} \,\widetilde{\psi}_f^*(\vect{k} + \vect{q})\, \widetilde{\psi}_i(\vect{k}) = \int d^3x \, e^{i \vect{q} \cdot \vect{x} }\, \psi_f^*( \vect{x} )\, \psi_i(\vect{x}) \,,
    \label{eq:wf_form_fac}
\end{align}
where $\overline{\left| \mathcal{M} \right|^2}$ is the spin averaged matrix element squared and we have defined a crystal form factor $f_{i\rightarrow f}$, written in terms of both momentum and position space representations of the wave functions. 

The transition rate per target mass, $R_{i \rightarrow f}$, is then given by
\begin{align}
    \label{eq:rate}
    R_{i \rightarrow f} = \frac{1}{\rho_T} \frac{\rho_\chi}{m_\chi} \int d^3v f_\chi(\vect{v}) \,\overline{\Gamma}_{i \rightarrow f} \,,
\end{align}
where $\rho_T$ is the target density, $\rho_\chi = 0.4 \text{ GeV}/\text{cm}^{3}$ is the local DM density, and $f_\chi$ is taken to be a boosted Maxwell-Boltzmann distribution. The total rate, $R$, is then simply the sum over all possible transitions from initial to final states. Since the only $\vect{v}$ dependence in Eq.~\eqref{eq:rate} comes from the energy conserving delta function, we perform the $\vect{v}$ integral first and define $g(\vect{q}, \omega) = 2 \pi \int d^3v f_\chi(v) \delta(\omega - \omega_\vect{q})$. This integral can be evaluated analytically (see {\it e.g.}\ Refs.~\cite{Coskuner:2019odd,Trickle:2019nya,Coskuner:2021qxo}):
\begin{align}
    g(\vect{q}, \omega) &
    = \frac{2\pi^2 v_0^2}{N_0}\frac{1}{q} \left( e^{-v_-^2/v_0^2} - e^{-v_\text{esc}^2/v_0^2} \right) ,\\
    v_- & = \text{min} \left\{ \frac{1}{q} \left| \omega + \frac{q^2}{2m_\chi} + \vect{q} \cdot \vect{v}_e  \right| , v_{\text{esc}} \right\} ,\label{eq:v_minus}
\end{align}
where $\omega = E_f - E_i$ is the deposited energy, and $N_0$ is a normalization factor such that $\int d^3\vect{v} f_\chi(\vect{v}) = 1$. 
We take the DM velocity distribution parameters to be $v_0 = 230$ km$/$s , $v_\text{esc} = 600$ km$/$s, and $v_e = 240$ km$/$s. 
The total rate then becomes
\begin{align}
    R = \frac{2}{16 V m_e^2 m_\chi^3} \frac{\rho_\chi}{\rho_T}\sum_{i, f} \int \frac{d^3q}{(2\pi)^3} \overline{\left| \mathcal{M}(\vect{q}) \right|^2} \,g(\vect{q}, \omega) \left| f_{i \rightarrow f}(\vect{q}) \right|^2 \, .
\end{align}
Here we focus on simple DM models, such as the kinetically mixed dark photon or leptophilic scalar mediator models. In these models $\mathcal{M}(\vect{q})$ can be factorized as $\mathcal{M}(q) = \mathcal{M}(q_0) \mathcal{F}_\text{med}(q_0/q) \left(f_e/f_e^0 \right)$, where $\mathcal{F}_\text{med}(q_0/q) = 1$ for a heavy mediator and $\mathcal{F}_\text{med}(q_0/q) = (q_0/q)^2$ for a light mediator, and $f_e/f_e^0$ is a screening factor discussed in more detail in Sec.~\ref{subsec:in-med}. As in previous works, we choose the reference momentum transfer to be $q_0 = \alpha m_e$. We can then finally write the rate in terms of a reference cross section,
\begin{align}
    \overline{\sigma}_e = \frac{\mu_{\chi e}^2}{16\pi m_\chi^2 m_e^2} \overline{|\mathcal{M}(q_0)|^2} \, ,
\end{align}
and find
\begin{align}
    \label{eq:rate_simple}
    R = \frac{2 \pi \overline{\sigma}_e}{V \mu_{\chi e}^2 m_\chi} \frac{\rho_\chi}{\rho_T} \sum_{i, f} \int \frac{d^3q}{(2\pi)^3} \left( \frac{f_e}{f_e^0} \right)^2 \mathcal{F}_\text{med}^2 \,g(\vect{q}, \omega) \left| f_{i \rightarrow f}(\vect{q}) \right|^2 \, .
\end{align}
Another useful quantity is the \textit{binned} rate (the rate for energy deposition between $\omega$ and $\omega + \Delta \omega$), $\Delta R_\omega$, defined as
\begin{align}
    \Delta R_\omega = \frac{2 \pi \overline{\sigma}_e}{V \mu_{\chi e}^2 m_\chi} \frac{\rho_\chi}{\rho_T}\sum_{i, f}& \,  \theta(\omega + \Delta \omega - E_f + E_i) 
    \,\theta(\omega - E_f + E_i) \nonumber \\
& \times \int \frac{d^3q}{(2\pi)^3} \left( \frac{f_e}{f_e^0} \right)^2\,\mathcal{F}_\text{med}^2 \,g(\vect{q}, \omega) \left| f_{i \rightarrow f}(\vect{q}) \right|^2 .
    & 
    \label{eq:binned_rate_E}
\end{align}

\subsection{Valence to Conduction}
\label{subsec:vc}

\begin{figure}[t]
    \centering
    \includegraphics[width=\textwidth]{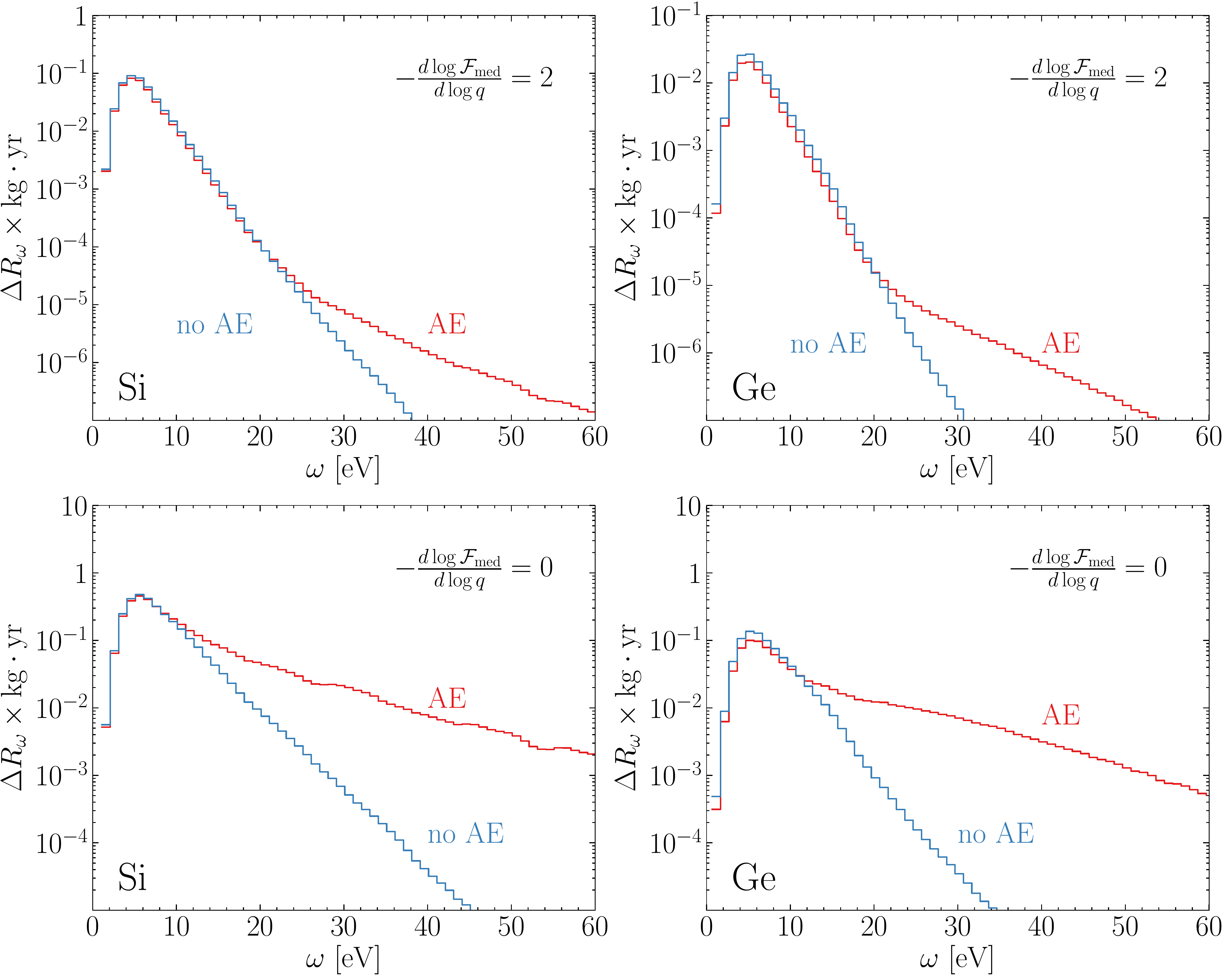}
    \caption{DM-electron scattering rate from valence to conduction bands binned in energy deposition (with $\Delta\omega=1$\,eV) for 1\,GeV DM, light (top row) and heavy (bottom row) mediators, assuming $\overline{\sigma}_e = 10^{-40}\; \text{cm}^2$, computed with vs.\ without AE reconstruction. Valence states included are the first four bands below the band gap, and conduction states included are all bands up to $E_\text{dft}=60$\,eV.}
    \label{fig:vc_diff_rate}
\end{figure}

We begin with valence to conduction band transitions. The initial (final) states are indexed by band number, $i (f)$, and Bloch momentum, $\vect{k}_i (\vect{k}_f)$ inside the 1BZ. The wave functions in Eq.~\eqref{eq:bloch_form} can be substituted into the crystal form factor in Eq.~\eqref{eq:wf_form_fac},
\begin{align}
    f_{i, \vect{k}_i \rightarrow f, \vect{k}_f} & = \frac{1}{V} \int d^3x \,e^{i \left( \vect{k}_i - \vect{k}_f + \vect{q} \right) \cdot \vect{x}}\, u^*_{f, \vect{k}_f}(\vect{x}) \,u_{i, \vect{k}_i}(\vect{x}) \nonumber \\
                                              & = \sum_\vect{G} \delta_{\vect{q},\, \vect{k}_f - \vect{k}_i + \vect{G}} \,\frac{1}{\Omega}\, \int_\text{cell} d^3x \,e^{i \vect{G} \cdot \vect{x}} \,u^*_{f, \vect{k}_f}(\vect{x}) \,u_{i, \vect{k}_i}(\vect{x}) \,,
                                \label{eq:f_val_to_cond}
\end{align}
where the integral is over the primitive cell with volume $\Omega$, and we have used the identity $\sum_\vect{r} e^{i \vect{q} \cdot \vect{r}} = N \sum_\vect{G} \delta_{\vect{q}, \vect{G}}$. The total rate in Eq.~\eqref{eq:rate_simple} is then
\begin{align}
    R = \frac{2 \pi \overline{\sigma}_e}{\mu_{\chi e}^2 m_\chi} \frac{\rho_\chi}{\rho_T} \sum_{i = 1}^{N_\text{v}} \sum_{f = 1}^{N_\text{c}} & \int_\text{1BZ} \frac{d^3k_i}{(2\pi)^3} \frac{d^3k_f}{(2\pi)^3} \sum_\vect{G} \left( \frac{f_e}{f_e^0} \right)^2 \mathcal{F}_\text{med}^2 \, g(\vect{q}, \omega) \nonumber \\
                                                                                                                                                       & \times \left| \frac{1}{\Omega} \int_\text{cell} d^3x \, e^{i \vect{G} \cdot \vect{x}} \,u^*_{f, \vect{k}_f}(\vect{x}) \,u_{i,\vect{k}_i}(\vect{x}) \right|^2 ,
      \label{eq:rate_val_to_cond}
\end{align}
where $\vect{q} = \vect{k}_f - \vect{k}_i + \vect{G}$, $N_{\text{v}(\text{c})}$ is the number of valence (conduction) bands. This is identical to the rate formulae derived in~\cite{Griffin:2019mvc, Trickle:2019nya, Essig:2015cda} but written in terms of the periodic Bloch functions, $u_{i,\vect{k}}(\vect{x})$, instead of their Fourier transformed components, $\widetilde{u}_{i,\vect{k},\vect{G}}$, similar to Ref.~\cite{Liang:2018bdb}. Numerically the position space form is superior since the integral over the primitive cell can be computed by Fast Fourier Transform. This reduces the computational complexity from $\mathcal{O}(N_G^2)$ to $\mathcal{O}(N_G \log{N_G})$, where $N_G$ is the number of $\vect{G}$ points, {\it i.e.}\ the number of Fourier components in the expansion of $\widetilde{u}_{i,\vect{k}}$ in Eq.~\eqref{eq:bloch_fourier}.

In Fig.~\ref{fig:vc_diff_rate} we show the scattering rate from valence to conduction transitions binned in energy deposition, defined in Eq.~\eqref{eq:binned_rate_E}, for a 1\,GeV DM. The main difference between the calculation performed here and in previous works is the effect of the AE reconstruction, as discussed in Sec.~\ref{subsubsec:ae}. For the case of DM with a heavy mediator, the rate, even with experimental thresholds as low as $\sim10$ eV, is significantly enhanced relative to previous work. The AE reconstruction plays less of a role in the light mediator case since the transition rate is dominated by small momentum transfers. However, at high thresholds, where only larger momentum components can contribute, the AE reconstruction can still significantly boost the scattering rate by fully including the contributions neglected in the pseudo wave functions. 

Since most earlier works computing DM-electron scattering include only valence to conduction transitions, it is useful to understand for which DM masses these are the only kinematically allowed transitions. If $\omega < E_\text{g} - E_\text{max}^\text{core}$, where $E_\text{max}^\text{core}$ is the maximum energy of the core states, then the core states cannot contribute; if $\omega < E_\text{dft}$ the free states are not available. Therefore if $\omega < \text{min}\{ E_\text{dft}, E_\text{g} - E_\text{max}^\text{core} \}$ only the valence to conduction transitions are allowed, which can be related to a DM mass via $\omega_\text{max}(m_\chi) < \text{min}\{ E_\text{dft}, E_\text{g} - E_\text{max}^\text{core} \}$, where
\begin{align}
    \omega_\text{max}(m_\chi) = \frac{1}{2} m_\chi v_\text{max}^2 = 3.9 \text{ eV} \left( \frac{m_\chi}{\text{MeV}} \right) \left( \frac{v_\text{max}}{840 \text{ km}/\text{s}} \right)^2 \, ,
    \label{eq:w_max}
\end{align}
with $v_\text{max} = v_e + v_\text{esc}$, the maximum incoming DM velocity. For Si (Ge), $E_\text{max}^\text{core} = -116$\,eV ($-28$\,eV), this corresponds to 
\begin{align}
    m_\chi <
    \begin{cases}
        15.2 \text{ MeV} & (\text{Si}) \,,\\
        7.8 \text{ MeV} & (\text{Ge}) \, .
    \end{cases}
\end{align}
Requiring that $\omega_\text{max} > E_\text{g}$, where $E_\text{g}$ is the band gap, sets a lower bound on the minimum detectable mass, $m_\chi^\text{min}$,
\begin{align}
    m_\chi^\text{min} = \frac{2E_\text{g}}{v_\text{max}^2} = 0.25 \text{ MeV} \left( \frac{E_\text{g}}{\text{eV}} \right) \left( \frac{840 \text{ km}/\text{s}}{v_\text{max}} \right)^2 \, .
\end{align}
For Si (Ge), with a band gap of 1.11 (0.67) eV, $m_\chi^\text{min}$ is $0.28$ ($0.17$) MeV.
Lastly, we remark that for DM interactions characterized by higher-dimensional operators (not considered in this work), the scattering rate scales with higher powers of $q$ and therefore is even more sensitive to AE reconstruction (and also $\mathrm{c} \rightarrow \mathrm{c}$ contributions discussed below in Sec.~\ref{subsec:cc}), which must be included in the analysis.

\subsection{Valence to Free}
\label{subsec:vf}

\begin{figure}[t]
    \centering
    \includegraphics[width=\textwidth]{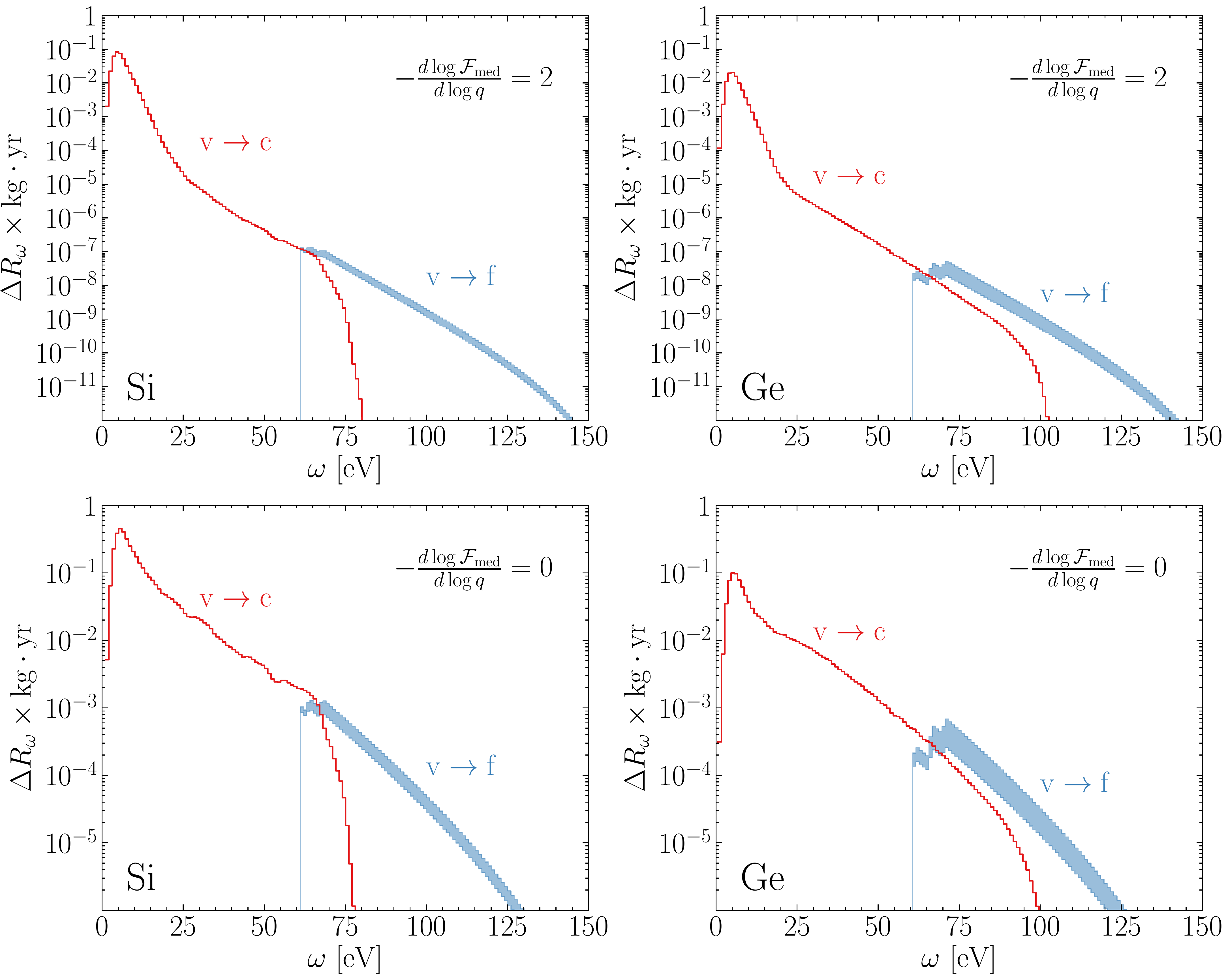}
    \caption{DM-electron scattering rate from valence to conduction (v$\to$c) bands and from valence bands to free states (v$\to$f) binned in energy deposition (with $\Delta\omega=1$\,eV) for 1\,GeV DM, light (top row) and heavy (bottom row) mediators, assuming $\overline{\sigma}_e = 10^{-40}\; \text{cm}^2$. The upper edge of the shaded region corresponds to using $Z_\text{eff}$ from Eq.~\eqref{eq:z_eff_binding_approx}, while the bottom edge corresponds to $Z_\text{eff} = 1$.}
    \label{fig:vf_diff_rate}
\end{figure}

For valence to free transitions the initial states are identical to those from Sec.~\ref{subsec:vc}, labeled by band number $i$ and Bloch momentum, $\vect{k}_i$. The final state wave functions are simple plane waves given by Eq.~\eqref{eq:free_elec_wf}, labeled by a momentum $\vect{k}_f$ in the 1BZ with the bands labeled by $\vect{G}$. 
We can therefore directly substitute Eq.~\eqref{eq:u_free} into Eq.~\eqref{eq:f_val_to_cond} derived in the previous subsection, and obtain the crystal form factor:
\begin{align}
 f_{i, \vect{k}_i \rightarrow \vect{G}_f,\vect{k}_f } &=
 \sum_\vect{G} \delta_{\vect{q},\, \vect{k}_f - \vect{k}_i + \vect{G}} \;\frac{1}{\Omega}\, \int_\text{cell} d^3x \,e^{i (\vect{G}-\vect{G}_f) \cdot \vect{x}}  \,u_{i, \vect{k}_i}(\vect{x}) \nonumber\\
 &= \sum_\vect{G} \delta_{\vect{q},\, \vect{k}_f - \vect{k}_i + \vect{G}} \,\widetilde{u}_{i,\vect{k}_i, \vect{G}_f - \vect{G}} \,,
\end{align}
where $\widetilde{u}_{i,\vect{k}_i,\vect{G}}$ are the Fourier components of the Bloch wave functions defined in Eq.~\eqref{eq:bloch_fourier}. 
Incorporating the Fermi factor correction discussed in Sec.~\ref{subsec:pw_approx}, we find the rate in Eq.~\eqref{eq:rate_simple} is given by
\begin{equation}
    R  = \frac{2 \pi \overline{\sigma}_e}{\mu_{\chi e}^2 m_\chi} \frac{\rho_\chi}{\rho_T} \sum_{i = 1}^{N_\text{v}} \sum_{\vect{G}_f} \int_\text{1BZ} \frac{d^3k_i}{(2\pi)^3} \frac{d^3k_f}{(2\pi)^3} F(\nu_{i,\vect{k}_i}) \sum_\vect{G} \left( \frac{f_e}{f_e^0} \right)^2  \mathcal{F}_\text{med}^2 \, g(\vect{q}, \omega) \left| \widetilde{u}_{i,\vect{k}_i, \vect{G}_f - \vect{G}}\right|^2 ,
\end{equation}
where
\begin{equation}
\omega  \equiv \frac{\left| \vect{k}_f + \vect{G}_f \right|^2}{2 m_e} - E_{i,\vect{k}_i} \,,\qquad
\nu_{i,\vect{k}_i} = \nu(Z_\text{eff}^{i, \vect{k}_i},\, \omega + E_{i,\vect{k}_i}) \,.
\end{equation}
With a change of variables, $\vect{G}' = \vect{G}_f - \vect{G}$ and defining $\vect{k}' \equiv \vect{k}_f + \vect{G}_f$ (and then dropping the prime for simplicity), the rate becomes
\begin{align}
    R & = \frac{2 \pi \overline{\sigma}_e}{\mu_{\chi e}^2 m_\chi} \frac{\rho_\chi}{\rho_T} \sum_{i = 1}^{N_\text{v}} \int_\text{1BZ} \frac{d^3k_i}{(2\pi)^3} F(\nu_{i,\vect{k}_i}) \sum_{\vect{G}} \left| \widetilde{u}_{i,\vect{k}_i,\vect{G}}\right|^2 \int \frac{d^3k}{(2\pi)^3}  \left( \frac{f_e}{f_e^0} \right)^2 \mathcal{F}_\text{med}^2 \, g(\vect{q}, \omega) \, .
\end{align}
where $\vect{q} = \vect{k} - \vect{k}_i - \vect{G}$.

In Fig.~\ref{fig:vf_diff_rate} we compare the binned rate from the valence to conduction (v$\to$c) calculation in the previous subsection to the valence to free (v$\to$f) one performed here, again for a 1\,GeV DM. We see that for large $\omega$, where the v$\to$c calculation is limited by the number of conduction bands included, the v$\to$f calculation extrapolates the results to higher $\omega$ as expected. There is some uncertainty due to the choice of the effective charge parameters, which is why the results are shown in bands. The lower edge corresponds to the conservative choice of $Z_\text{eff}^{i, \vect{k}_i} = 1$ for all $i, \vect{k}_i$, and the upper edge corresponds to the value set by the binding energy, Eq.~\eqref{eq:z_eff_binding_approx} with $E_B=-E_{i,\vect{k}_i}$. We find that the conservative choice $Z_\text{eff}^{i, \vect{k}_i} = 1$ is a better match to the edge for the v$\to$c calculation, and will use this in our final projections in Sec.~\ref{sec:projected_constraints}. Note that as the threshold increases, the effect of v$\to$f transitions becomes more important, and for a heavy mediator non-negligible constraints can be placed even with $\mathcal{O}(100)$\,eV energy thresholds.

\subsection{Core to Conduction}
\label{subsec:cc}

\begin{figure}[t]
    \centering
    \includegraphics[width=\textwidth]{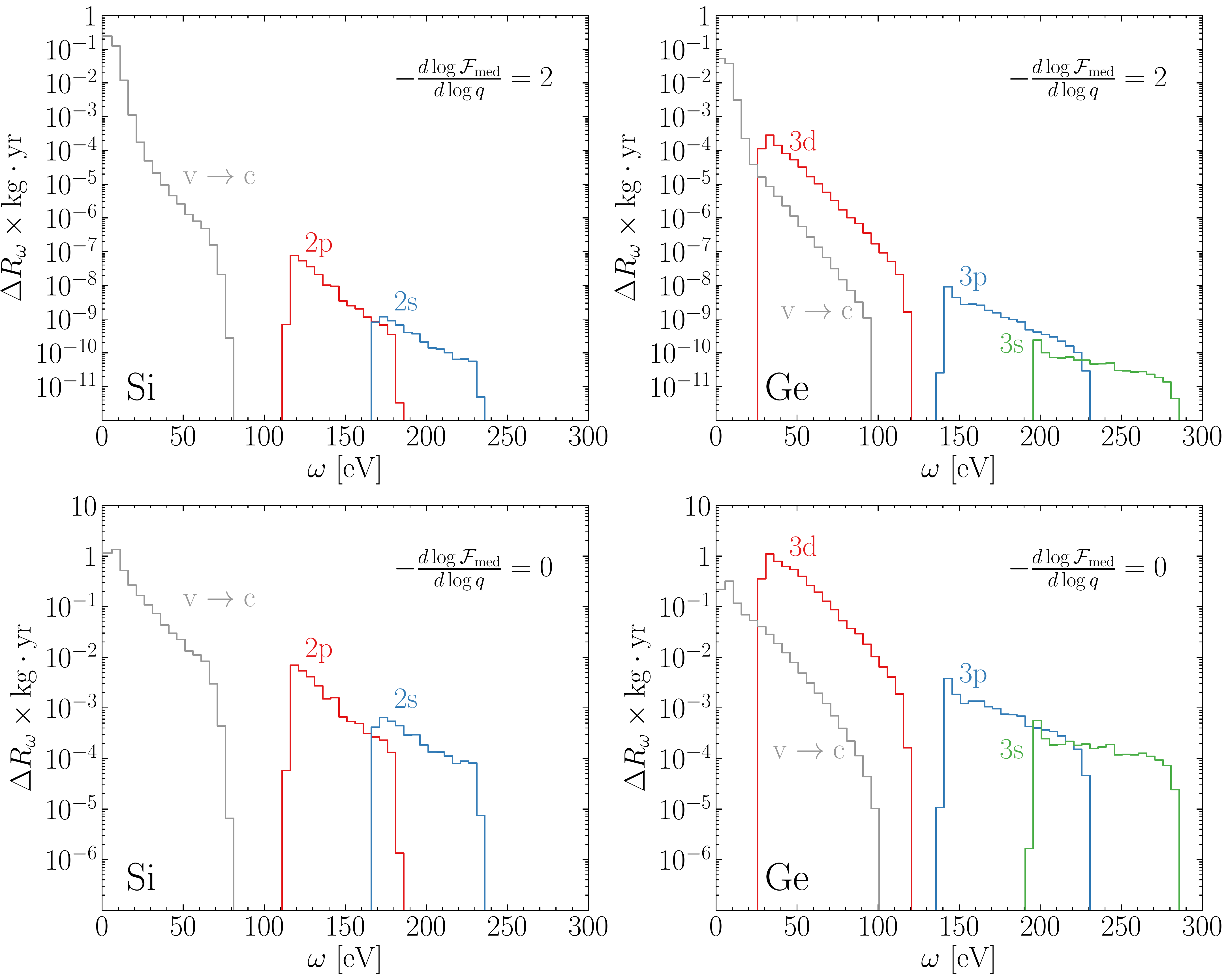}
    \caption{DM-electron scattering rate from core states to conduction bands binned in energy deposition (with $\Delta\omega=5$\,eV) for 1\,GeV DM, light (top row) and heavy (bottom row) mediators, assuming $\overline{\sigma}_e = 10^{-40}\; \text{cm}^2$. The core states are labelled by the corresponding atomic orbitals, and the conduction states up to $E_\text{dft}=60$\,eV are included. For comparison we also show the v$\to$c contribution (after AE reconstruction) from Fig.~\ref{fig:vc_diff_rate} in gray.}
    \label{fig:cc_diff_rate}
\end{figure}

We now turn to core to conduction transitions. The initial core states are indexed by $\kappa$, the atom in the primitive cell, the usual atomic quantum numbers, $n, l, m$, and the Bloch momentum, $\vect{k}_i$. The final states are the DFT computed conduction states. The crystal form factor is simply obtained from Eq.~\eqref{eq:f_val_to_cond} by substituting $u_{i,\vect{k}_i}\to u_{\kappa nlm,\vect{k}_i}$:
\begin{equation}
f_{\kappa nlm, \vect{k}_i \rightarrow f, \vect{k}_f}  =\sum_\vect{G} \delta_{\vect{q},\, \vect{k}_f - \vect{k}_i + \vect{G}} \,\frac{1}{\Omega}\, \int_\text{cell} d^3x \,e^{i \vect{G} \cdot \vect{x}} \,u^*_{f, \vect{k}_f}(\vect{x}) \,u_{\kappa nlm, \vect{k}_i}(\vect{x}) \,,
\label{eq:f_c2c}
\end{equation}
The total scattering rate is then
\begin{align}
    R = \frac{2\pi\overline{\sigma}_e}{\mu_{\chi e}^2 m_\chi} \frac{\rho_\chi}{\rho_T} \sum_{\kappa = 1}^{N_\text{a}} \sum_{n = 1}^{N_\text{p}^\kappa} & \sum_{l = 0}^{n-1} \sum_{m = -l}^{l} \sum_{f = 1}^{N_\text{c}} \int_\text{1BZ} \frac{d^3k_i}{(2\pi)^3} \frac{d^3k_f}{(2\pi)^3} \sum_\vect{G} \left( \frac{f_e}{f_e^0} \right)^2 \mathcal{F}_\text{med}^2\, g(\vect{q}, \omega) \nonumber \\  
                                                                                                                                                       & \times \left| \frac{1}{\Omega} \int_\text{cell} d^3x \, e^{i \vect{G} \cdot \vect{x}} u_{f\vect{k}_f}^*(\vect{x}) \,u_{\kappa nlm\vect{k}_i}(\vect{x}) \right|^2 ,
\end{align}
where $N_\text{a}$ is the number of atoms in the primitive cell, $N_\text{p}^\kappa$ is the largest principal quantum number for atom $\kappa$, and $\omega = E_{f,\vect{k}_f} - E_{\kappa nl}$. 
The core wave functions $u_{\kappa nlm,\vect{k}_i}(\vect{x})$ are given by Eq.~\eqref{eq:u_jnlmk}, and involves a sum over primitive cells. 
Since the integral in Eq.~\eqref{eq:f_c2c} is just over one primitive cell, only the atoms in this and neighboring cells can have a significant contribution. 
In other words, the sum over $\vect{r}$ converges very quickly due to the localized nature of atomic wave functions. We therefore restrict $\vect{r}$ to be summed over only the $3\times 3\times 3$ nearest cells.

\begin{figure}[t]
    \centering
    \includegraphics[width=\textwidth]{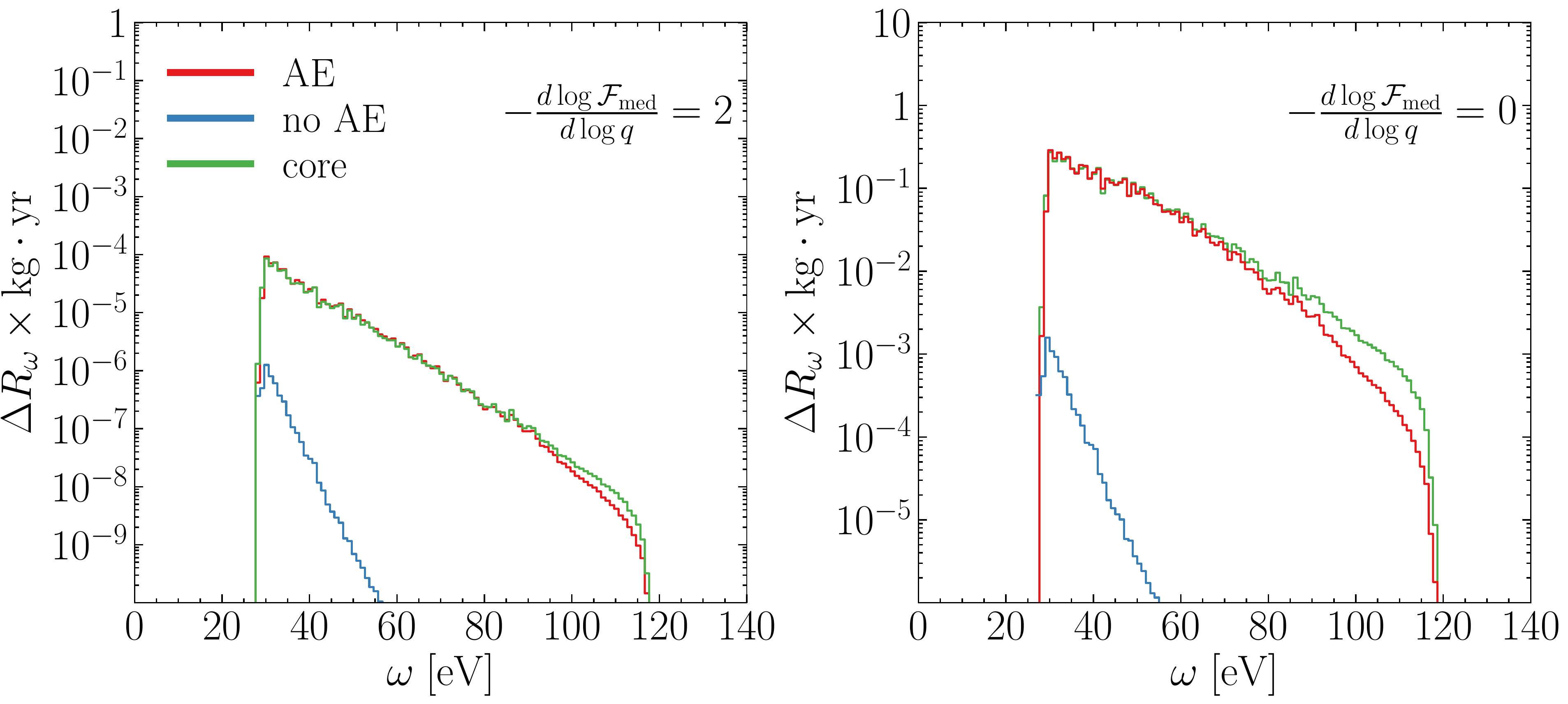}
    \caption{Contribution to the DM-electron scattering rate binned in energy deposition (with $\Delta\omega=1$\,eV) from 3d electrons to conduction bands in Ge, for 1\,GeV DM, light (left) and heavy (right) mediators, assuming $\overline{\sigma}_e = 10^{-40}\; \text{cm}^2$. The three curves in each panel are computed using DFT with and without AE reconstruction, and using the semi-analytic core wave functions.}
    \label{fig:Ge_3d_focus}
\end{figure}

The contribution of core to conduction (c$\to$c) transitions to the binned rate, for $m_\chi = 1$\,GeV, can be seen in Fig.~\ref{fig:cc_diff_rate}. In most cases the v$\to$c transitions are dominant compared to the c$\to$c, but there are two main scenarios where this is not true. First, when the experimental threshold is raised; this excludes the v$\to$c transitions and causes the c$\to$c contribution to be dominant. For example, consider a Si detector and a DM model with a heavy mediator (bottom left panel of Fig.~\ref{fig:cc_diff_rate}). If the experimental threshold is $\sim 50$\,eV the c$\to$c contribution from the 2p states in Si gives the dominant contribution. 
Second, for a Ge target, and a DM model with a heavy mediator, the 3d states dominate the rate even at the lowest experimental threshold. To understand this in more detail we present Fig.~\ref{fig:Ge_3d_focus} which compares the binned rate taking different modeling approaches for the 3d states in Ge. We see that the large momentum components of the wave function, recovered only after AE reconstruction in the DFT calculation, dominate the rate, which explains why previous works have underestimated the importance of 3d electrons. 
Meanwhile, we see explicitly at the scattering rate level that the semi-analytic approach accurately reproduces the DFT calculation at low $\omega$, and extends the latter beyond its cutoff at high $\omega$, consistent with the observation at the wave function level in Fig.~\ref{fig:binned_core_wf}.

\subsection{Core to Free}
\label{subsec:cf}

\begin{figure}[t]
    \centering
    \includegraphics[width=\textwidth]{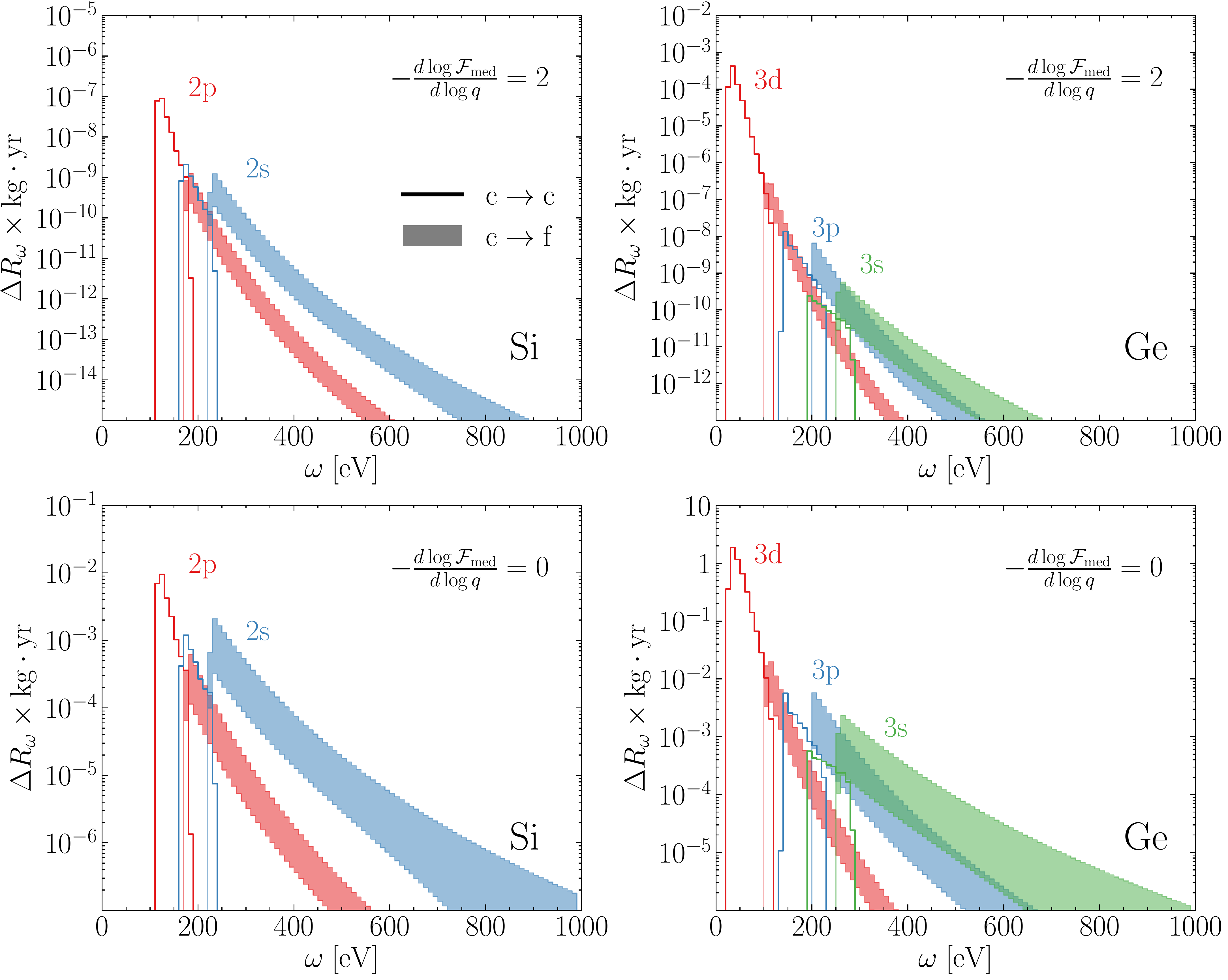}
    \caption{DM-electron scattering rate from core states to conduction bands (c$\to$c) and to free states (c$\to$f) binned in energy deposition (with $\Delta\omega=10$\,eV) for 1\,GeV DM, light (top row) and heavy (bottom row) mediators, assuming $\overline{\sigma}_e = 10^{-40}\; \text{cm}^2$.  As in the v$\to$f calculation in Fig.~\ref{fig:vf_diff_rate}, the upper edge of the shaded bands corresponds to $Z_\text{eff}$ from Eq.~\eqref{eq:z_eff_binding_approx}, and the lower edge corresponds to $Z_\text{eff} = 1$.}
    \label{fig:cf_diff_rate}
\end{figure}

The last transition type we consider involves a core electron initial state and a free electron final state.
The crystal form factor is most easily obtained by substituting Eqs.~\eqref{eq:core_elec_wf} and \eqref{eq:free_elec_wf} into its definition, Eq.~\eqref{eq:wf_form_fac}:
\begin{align}
    f_{\kappa nlm,\vect{k}_i \rightarrow \vect{G}_f, \vect{k}_f} & = \frac{1}{\sqrt{N V}} \sum_\vect{r} e^{i\vect{k}_i \cdot (\vect{r}+\vect{x}_\kappa)} \int d^3x\, e^{i (\vect{q} - \vect{k}_f - \vect{G}_f) \cdot \vect{x}} \,\psi_{\kappa nlm}^\text{atom}(\vect{x} - \vect{r} - \vect{x}_\kappa) \nonumber \\
                                   & = \frac{1}{\sqrt{NV}} \,e^{i(\vect{k}_i+\vect{q} - \vect{k}_f - \vect{G}_f) \cdot \vect{x}_\kappa} \sum_\vect{r} e^{i (\vect{q} - \vect{k}_f + \vect{k}_i) \cdot \vect{r}} \int d^3x \,e^{i (\vect{q} - \vect{k}_f - \vect{G}_f) \cdot \vect{x}} \,\psi_{\kappa nlm}^\text{atom}(\vect{x}) \nonumber \\
                                   & = \frac{1}{\sqrt{\Omega}} \,e^{i(\vect{k}_i+\vect{q} - \vect{k}_f - \vect{G}_f) \cdot \vect{x}_\kappa} \sum_\vect{G} \delta_{\vect{q} - \vect{k}_f + \vect{k}_i, \vect{G}}\, \widetilde{\psi}_{\kappa nlm}^\text{atom}(-\vect{k}_i + \vect{G} - \vect{G}_f) \,,
\end{align}
where the Fourier transform of the RHF Slater type orbital (STO) core wave functions, given in Eq.~\eqref{eq:sto_wf}, are known analytically \cite{Belkic1989}:
\begin{align}
    \widetilde{\psi}_\text{STO}(\vect{q}; Z, n, l, m) & = \int d^3x \, e^{i \vect{q} \cdot \vect{x}} R_\text{STO}(x; Z, n) \,Y_l^m(\hat{\vect{x}}) \equiv \chi_\text{STO}(q; Z, n, l) \,Y_l^m(\hat{\vect{q}}) \,,\\
    \chi_\text{STO}(q; Z, n) & = 4 \pi N (n - l)! (2 Z)^{n} \left( \frac{i a_0 q}{Z} \right)^l \sum_{s = 0}^{ \lfloor (n - l)/2 \rfloor} \frac{\omega_s^{nl}}{\left( (a_0 q)^2 + Z^2 \right)^{n - s + 1}} \,,\\
    \omega_s^{nl} & = \left(-4 Z^2\right)^{-s} \frac{(n - s)!}{s! (n - l - 2s)!} \,.
\end{align}
The direct detection rate is then
\begin{align}
    R = &\; \frac{2\pi\overline{\sigma}_e}{\mu_{\chi e}^2 m_\chi} \frac{\rho_\chi}{\rho_T \Omega} \sum_{\kappa = 1}^{N_\text{a}} \sum_{n = 1}^{N_\text{p}^\kappa} \sum_{l = 0}^{n-1} \sum_{m = -l}^{l} \int_\text{1BZ} \frac{d^3k_i}{(2\pi)^3} \frac{d^3k_f}{(2\pi)^3}\nonumber \\                                                                                                                                                               & \times  \sum_{\vect{G}_f} \sum_\vect{G} F (\nu_{\kappa nl})  \left( \frac{f_e}{f_e^0} \right)^2 \mathcal{F}_\text{med}^2\, g(\vect{q}, \omega) \left| \widetilde{\psi}_{\kappa nlm}^\text{atom}\left( -\vect{k}_i + \vect{G} - \vect{G}_f \right) \right|^2 ,
\end{align}
where $\vect{q} = \vect{k}_f - \vect{k}_i + \vect{G}$, and $\nu_{\kappa nl}=\nu(Z_\text{eff}^{\kappa nl},\, \omega + E_{\kappa nl})$. We can now shift the $\vect{G}_f$ variable, $\vect{G'} \equiv \vect{G}_f - \vect{G}$ and define $\vect{k} = \vect{k}_i + \vect{G}'$ and $\vect{k}' = \vect{k}_f + \vect{G}$. Therefore, $\vect{q} = \vect{k}' - \vect{k}$ and
\begin{align}
    R = \frac{2\pi\overline{\sigma}_e}{\mu_{\chi e}^2 m_\chi} \frac{\rho_\chi}{\rho_T \Omega} & \sum_{\kappa = 1}^{N_\text{a}} \sum_{n = 1}^{N_\text{p}^\kappa} \sum_{l = 0}^{n - 1} \sum_{m = -l}^{l} \int \frac{d^3k}{(2\pi)^3} \frac{d^3k'}{(2\pi)^3} \,F( \nu_{\kappa nl} ) \left( \frac{f_e}{f_e^0} \right)^2 \mathcal{F}_\text{med}^2\, g(\vect{q}, \omega) \left| \widetilde{\psi}^\text{atom}_{\kappa nlm}\left( \vect{k} \right) \right|^2 ,
\end{align}
which is the closest expression to the vacuum matrix element, with just the inclusion of the core wave functions acting as a form factor.

In Fig.~\ref{fig:cf_diff_rate}, we compare the binned rate from the core to conduction (c$\to$c) calculation to the core to free (c$\to$f) calculation and see a reasonable extrapolation to higher $\omega$. As with the transition region between v$\to$c and v$\to$f shown in Fig.~\ref{fig:vf_diff_rate}, we find $Z_\text{eff} = 1$ gives a better match between c$\to$c and c$\to$f. While the total number of electrons from these transitions is expected to be much less than lower energy transitions, this is the best available calculation for thresholds up to the kinematic limit of $\omega_\text{max}$. 

\subsection{In-medium Screening}
\label{subsec:in-med}

\begin{figure}[t]
	\centering
	\includegraphics[width=0.45\textwidth]{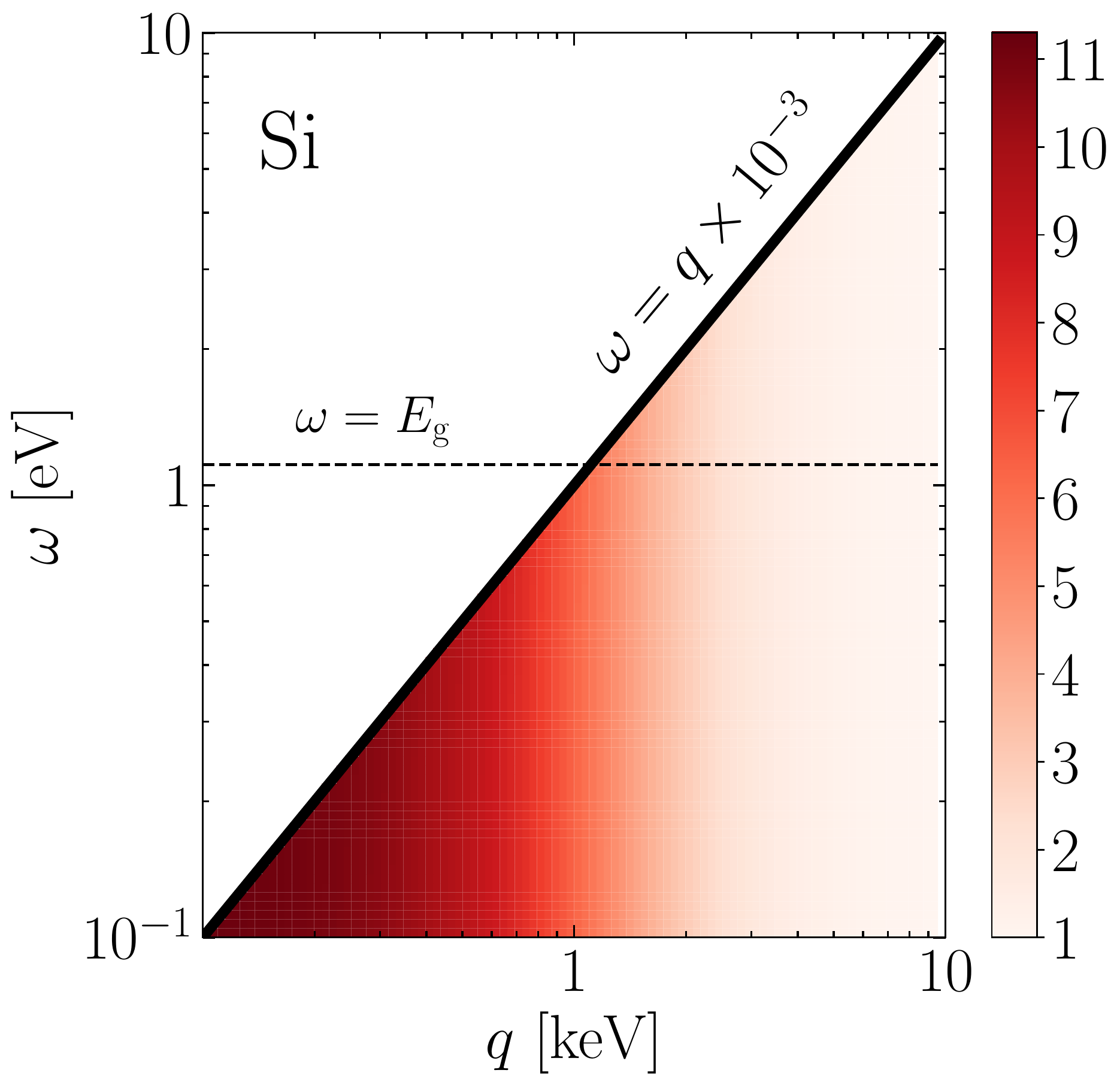}
	\includegraphics[width=0.45\textwidth]{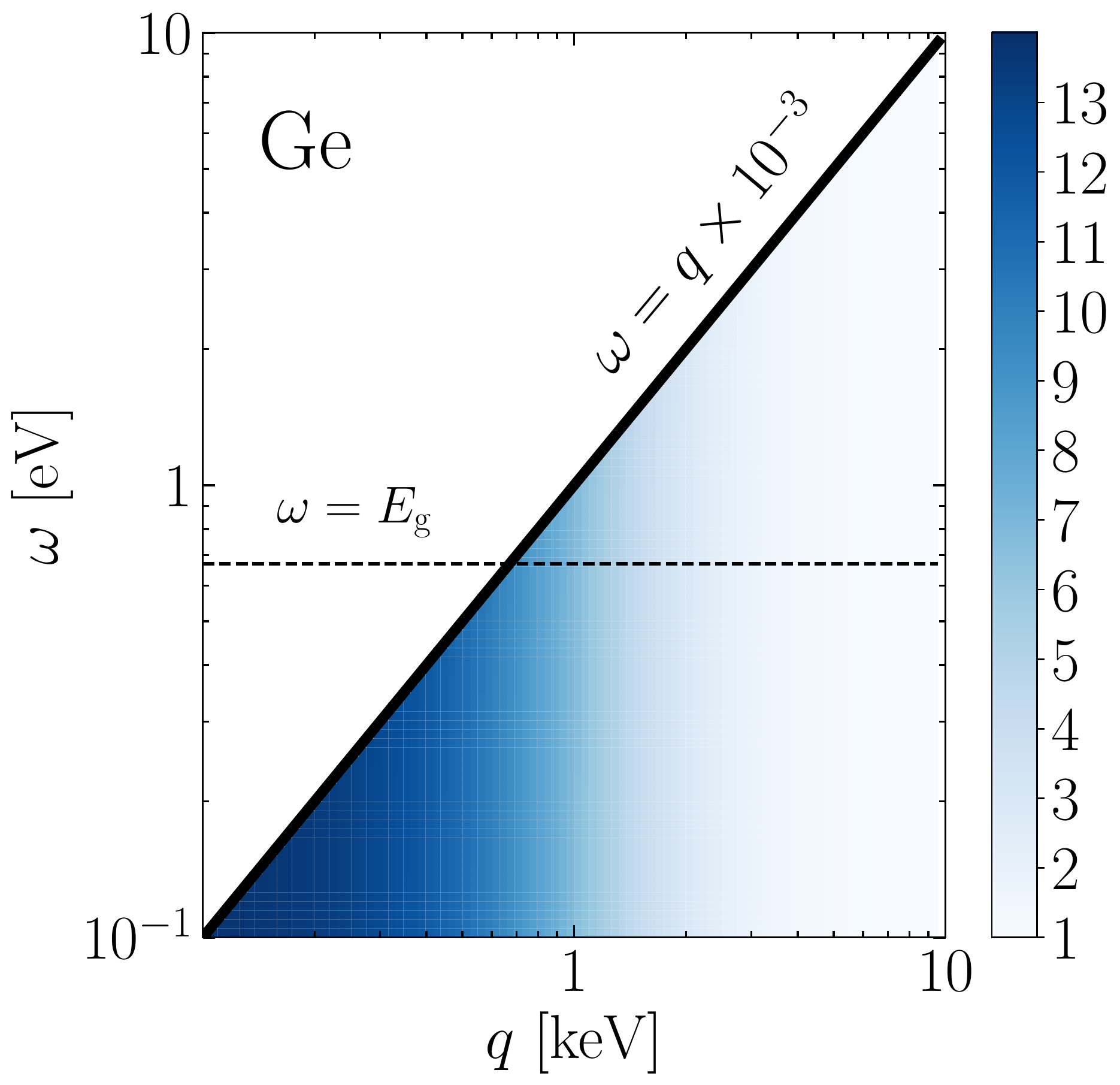}
	\caption{Dielectric function $\epsilon(q, \omega)$, given by Eq.~\eqref{eq:analytic_di} with the parameters in Table~\ref{tab:in_med}, of Si (left) and Ge (right) used to incorporate screening effects. The solid line indicates the edge of the kinematically accessible region $\omega \lesssim q v$. The dashed line is the band gap of the target. While the static dielectric can be $\mathcal{O}(10)$, in the kinematically allowed region $\epsilon(q, \omega)$ is an $\mathcal{O}(1)$ number, leading to an $\mathcal{O}(1)$ effect on the scattering rates when the latter are dominated by small $q, \omega$ transitions.}
	\label{fig:dielectrics}
\end{figure}

DM-electron interactions mediated by a dark photon or scalar are screened due to the in-medium mixing between the mediator and the photon. 
The relevance of screening has been recently emphasized in Ref.~\cite{Knapen:2021run}. 
The screening factor, $f_e/f_e^0$, is related to the longitudinal dielectric, $f_e/f_e^0 = (\hat{\vect{q}} \cdot \boldsymbol{\epsilon} \cdot \hat{\vect{q}})^{-1}$, where $\boldsymbol{\epsilon}$ is the dielectric tensor. 
It can be computed from in-medium loop diagrams or extracted from optical data. 
Here we model the dielectric of Si and Ge following Ref.~\cite{Cappellini1993}:
\begin{align}
    \epsilon(q, \omega) = 1 + \left[ \frac{1}{\epsilon_0 - 1} + \alpha \left( \frac{q}{q_\text{TF}} \right)^2 + \frac{q^4}{4m_e^2 \omega_p^2} - \left( \frac{\omega}{\omega_p} \right)^2  \right]^{-1} \, ,
    \label{eq:analytic_di}
\end{align}
and $\epsilon_{ij} = \epsilon(q, \omega) \,\delta_{ij}$. 
Here, $\epsilon_0 \equiv \epsilon(0, 0)$ is the static dielectric, $\alpha$ is a fitting parameter, $q_\text{TF}$ is the Thomas-Fermi momentum, and $\omega_p$ is the plasma frequency. The parameters used for Si and Ge are listed in Table~\ref{tab:in_med}, and we plot the dielectric as a function of $q, \omega$ in Fig.~\ref{fig:dielectrics}. 

\begin{table}[t]
    \begin{center}
        \begin{tabular}{c||c c c c}
            \hline
            Target & $\epsilon_0$ & $\alpha$ & $\omega_p$ $[\text{eV}]$ & $q_\text{TF}$ $[\text{keV}]$ \\
            \hline
            Si & 11.3 & 1.563 & 16.6 & 4.13 \\
            Ge & 14 & 1.563 & 15.2 & 3.99 \\
            \hline
        \end{tabular}
        \caption{Parameters used in the model of dielectric function, Eq.~\eqref{eq:analytic_di}, of Si and Ge from Ref.~\cite{Cappellini1993}, which accounts for in-medium screening effects on the transition rate.}
        \label{tab:in_med}
    \end{center}
\end{table}

\begin{figure}[t]
\centering
    \includegraphics[width=\textwidth]{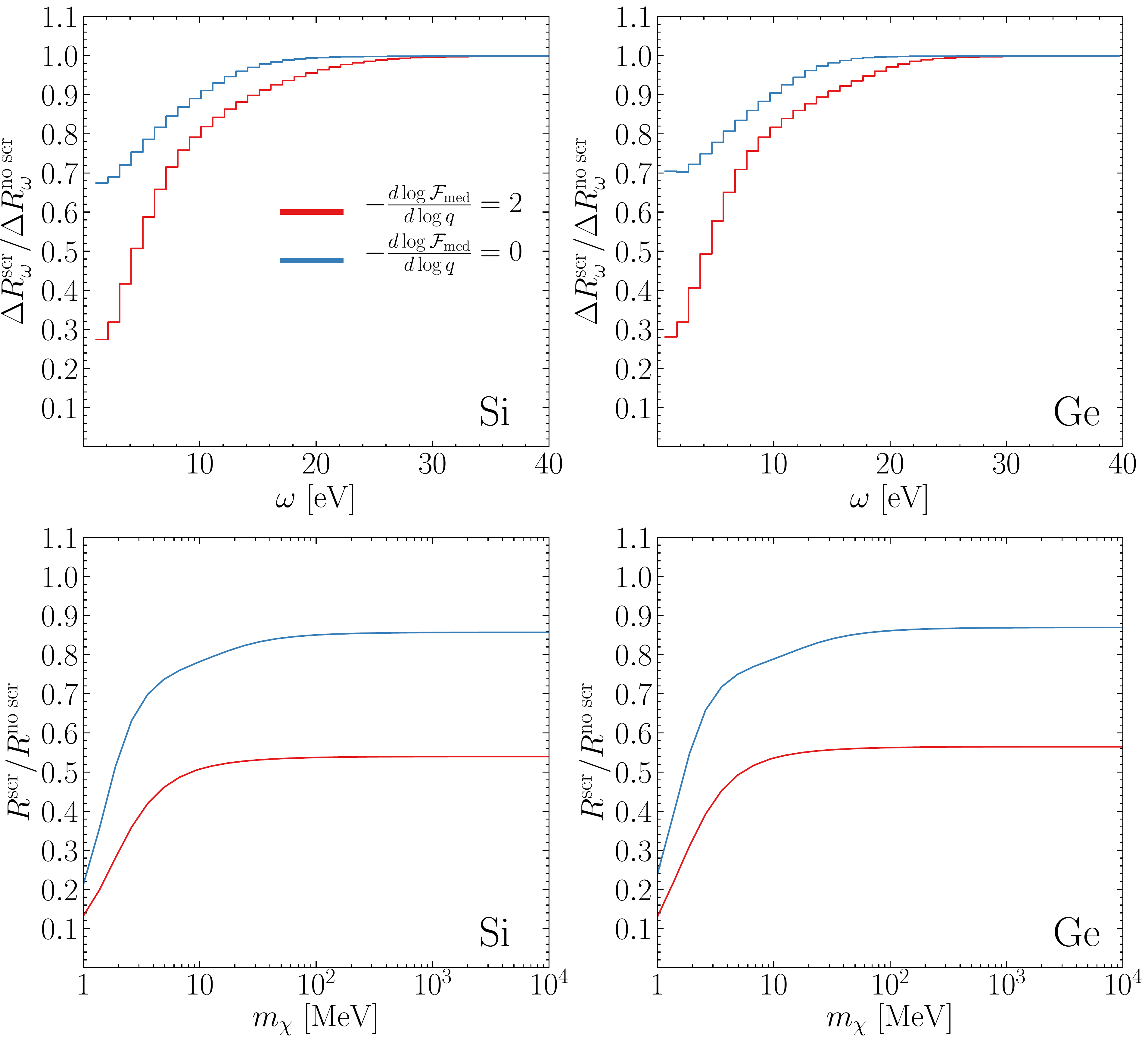}
    \caption{Effect of screening on the binned rate (top row, for 1\,GeV DM) and total rate (bottom row, as a function of $m_\chi$) from v$\to$c transitions for DM models with a light (red) and heavy (blue) mediator. The unscreened rate $R^\text{no scr}$ is obtained with $\epsilon=1$, and the screened rate $R^\text{scr}$ is obtained with the model of the dielectric function given in Eq.~\eqref{eq:analytic_di}.}
    \label{fig:scr_effects}
\end{figure}

Naively one might expect that the effect of the dielectric is to screen the rate by an $\mathcal{O}(100)$ factor due to the fact that the static dielectric, $\epsilon_0$, is $\mathcal{O}(10)$. However, this is only the value of the dielectric function at $q = \omega = 0$, while as $q \rightarrow \infty$ and $\omega \rightarrow \infty$ the dielectric approaches unity. Therefore, the effect of the dielectric crucially depends on the region of the kinematic phase space being probed. For a given energy deposition, $\omega$, the momentum transfer is limited to $q \gtrsim \omega / v$ where $v \sim 10^{-3}$ is the DM velocity. Therefore, the absolute minimum momentum transfer is $q_\text{min} \sim E_\text{g} / v \sim \mathcal{O}(\text{keV})$, for $\mathcal{O}(\text{eV})$ band gap targets. This is parametrically the same size as the Thomas-Fermi momentum $q_\text{TF}$, so the dielectric is expected to slightly deviate from one, which causes only an $\mathcal{O}(1)$ shift to the scattering rate, as seen in Fig.~\ref{fig:scr_effects}.

\section{Projected Sensitivity}
\label{sec:projected_constraints}

We now compile the results from the previous sections to compute the projected sensitivity. We also compare the relative importance of each transition type, and discuss differences between our results and previous calculations in the literature. When there are large differences, it is typically because of the inclusion of AE reconstruction and core states in the calculation. Since AE reconstruction and core states contribute predominantly at higher momentum transfer and energy deposition, we will find the largest differences typically occur for a massive mediator and higher detector threshold, where the effects in some cases can be more than an order of magnitude (especially for Ge). For the case of a massless mediator and lower detection threshold, the differences with previous literature are much smaller and mostly due to the inclusion of in-medium effects.

\begin{figure}[t]
    \centering
    \includegraphics[width=\textwidth]{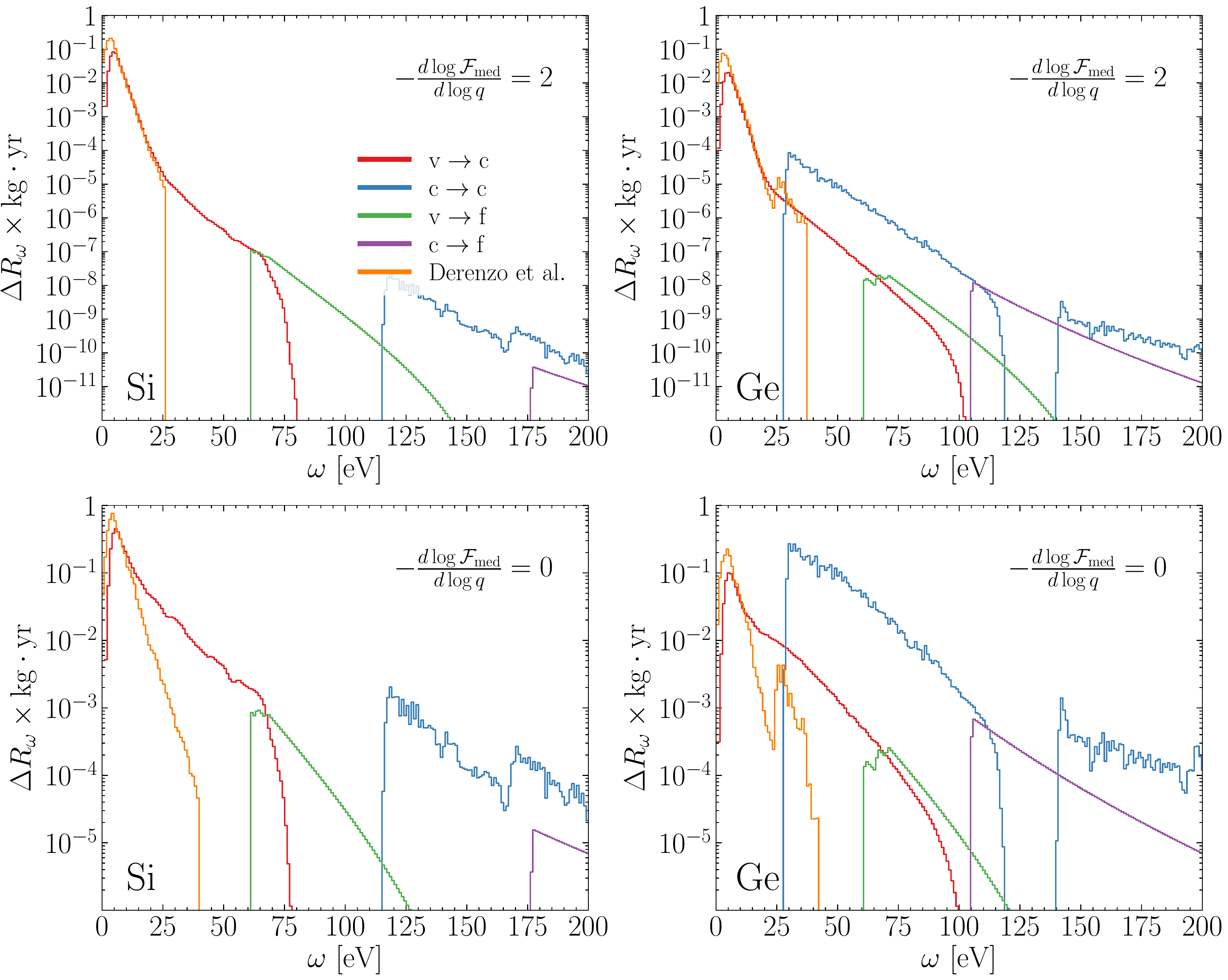}
    \caption{DM-electron scattering rate binned in energy deposition (with $\Delta\omega=1$\,eV)  for 1\,GeV DM, light (top row) and heavy (bottom row) mediators, from all four transition types: valence to conduction (v$\to$c), valence to free (v$\to$f), core to conduction (c$\to$c), and core to free (c$\to$f). We assume $\overline{\sigma}_e = 10^{-40}\; \text{cm}^2$, and take $Z_\text{eff} = 1$ for all effective charges in the Fermi factor. Note that the c$\to$c and c$\to$f transitions involve semi-analytic treatment of 2p (3d) states and below in Si (Ge), which has been validated with DFT calculations including AE reconstruction; see Fig.~\ref{fig:binned_core_wf}. We also overlay the binned rate from Ref.~\cite{Derenzo:2016fse} which computed the v$\to$c contribution using \texttt{QEdark} (treating 3d states in Ge as valence, without including AE reconstruction effects).}
    \label{fig:compiled_diff_rate}
\end{figure}

\begin{figure}[t]
    \centering
    \includegraphics[width=\textwidth]{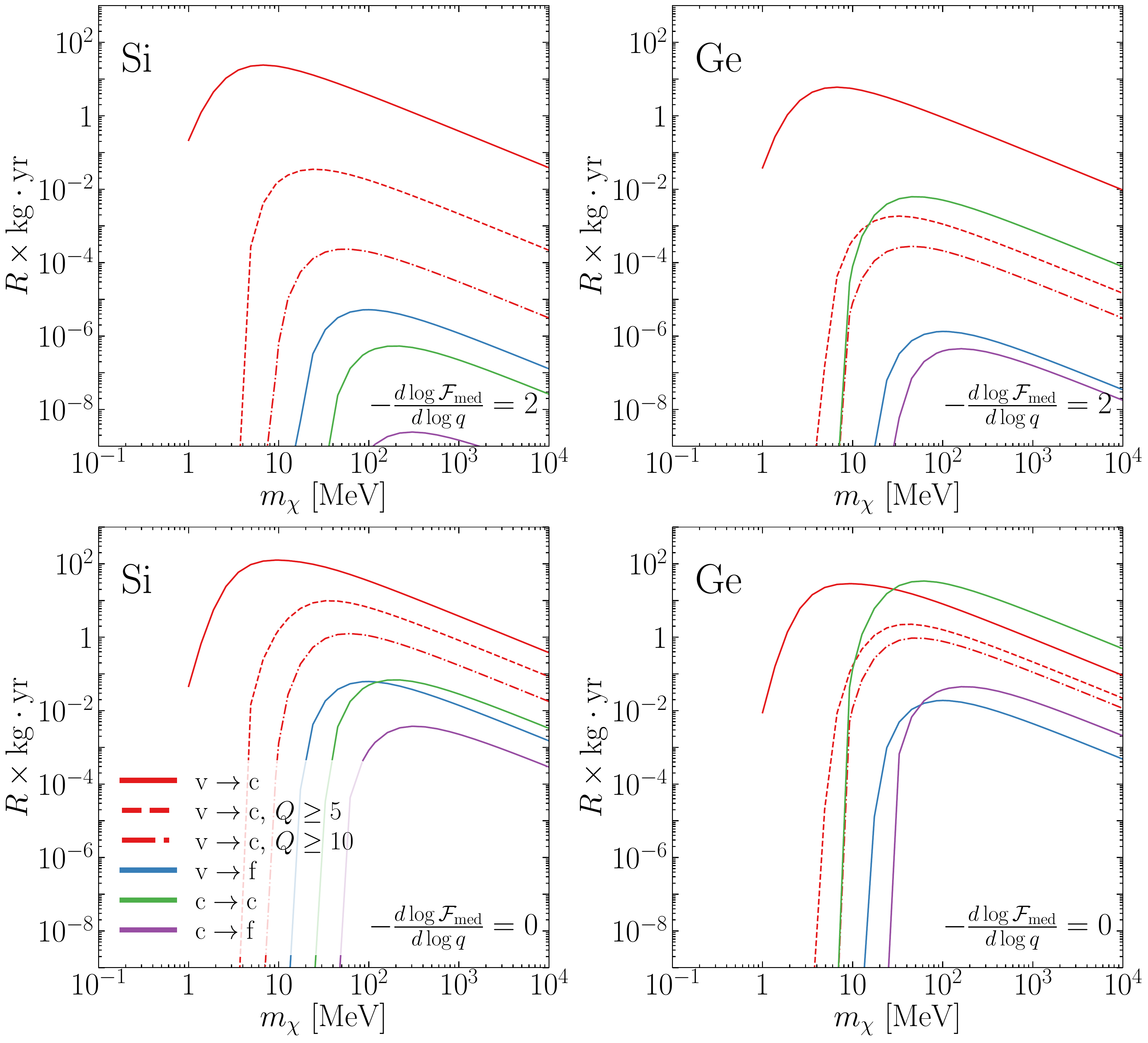}
    \caption{DM-electron scattering rate as a function of the DM mass, for light (top row) and heavy (bottom row) mediators, from all four transition types: valence to conduction (v$\to$c), valence to free (v$\to$f), core to conduction (c$\to$c), and core to free (c$\to$f). We assume $\overline{\sigma}_e = 10^{-40}\; \text{cm}^2$, take $Z_\text{eff} = 1$ for all effective charges in the Fermi factor, and show results for several threshold $Q$ values which significantly impact the v$\to$c contribution.}
    \label{fig:compiled_rate}
\end{figure}

In Fig.~\ref{fig:compiled_diff_rate} we show the contribution to the binned rate from each of the four transition types, for a 1\,GeV DM. 
We see that valence to conduction (v$\to$c) has a higher peak than the other three transition types, except for the Ge, heavy mediator case, where core to conduction (c$\to$c) has the highest peak. 
For comparison, Refs.~\cite{Essig:2015cda, Derenzo:2016fse} compute the valence to conduction rates with DFT, including also the 3d states in Ge, but without AE reconstruction. 
As expected, we find a lower rate at the lowest energy depositions due to the inclusion of in-medium screening, and a much higher rate at high $\omega$ due to AE reconstruction and inclusion of core states. 

The impact of these observations on the reach depends on the energy threshold. 
Assuming charge readout ({\it e.g.}\ via a CCD), the relevant quantity is the number of electron-hole pairs, $Q$, produced in an event. 
For an energy deposition $\omega$, this is given by
\begin{align}
Q = 1 + \left\lfloor \frac{\omega - E_\text{g}}{\varepsilon} \right\rfloor ,
\label{eq:Q_bin}
\end{align}
where the values for $\varepsilon$ are $3.6$ eV and $2.9$ eV for Si and Ge respectively. 
In Fig.~\ref{fig:compiled_rate}, we show the total rate as a function of the DM mass, for $Q\ge1, 5, 10$.
The threshold only affects the v$\to$ c rate, as the other three transition types involve energy depositions corresponding to $Q>10$, and are therefore always fully included. 
We see that for $Q\ge1$, the valence to conduction (v$\to$c) contribution dominates the total rate with the exception of the Ge, heavy mediator scenario, where core to conduction (c$\to$c) is dominant for $m_\chi\gtrsim 30\,$MeV. 
Higher thresholds significantly cut out v$\to$c contributions in all cases, and render c$\to$c more important for Ge, even in the light mediator scenario. 
For Si, on the other hand, the total rate is still dominated by v$\to$c because the core states are much deeper and contribute a lower rate. 
We also see that v$\to$f and c$\to$f contributions are subdominant in all cases.

Finally, we present the projected reach on the DM-electron reference cross section $\overline\sigma_e$ in Figs.~\ref{fig:reach_compare_Q1} and \ref{fig:reach_compare_Q10}, for $Q\ge1$ and $Q\ge10$, respectively. 
Our new calculation yields several important differences compared to the previous literature, and we discuss them in detail in the following subsection.

\subsection{Comparison With Previous Results}
\label{subsec:compare_to_prev}

\begin{figure}[t]
    \centering
    \includegraphics[width=\textwidth]{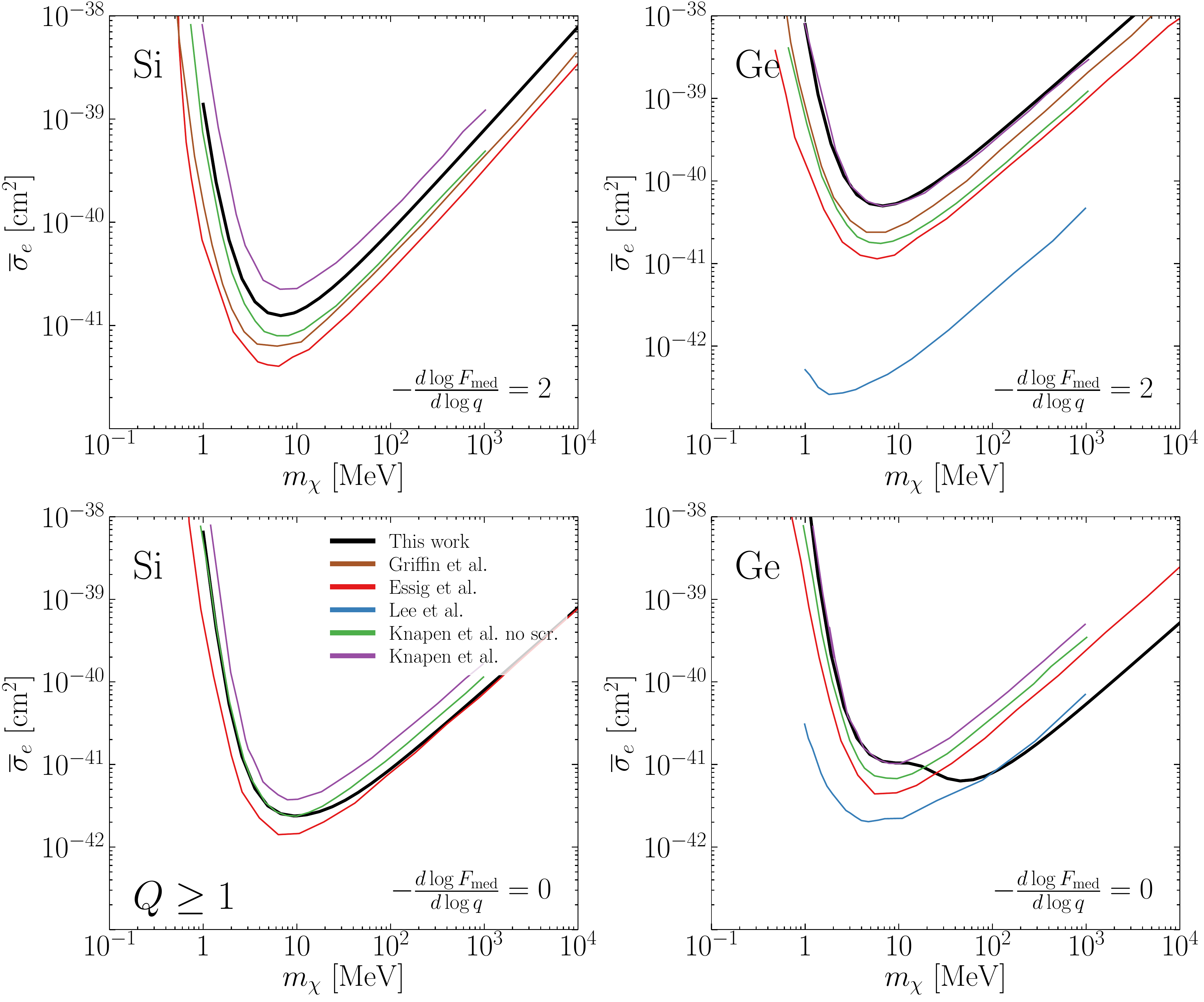}
    \caption{95\% C.L.\ exclusion reach (3 events) assuming 1 kg-year exposure, $Q \ge 1$, for light (top row) and heavy (bottom row) mediators. The results shown are from this work, Griffin et al.~\cite{Griffin:2019mvc}, Essig et al.~\cite{Essig:2015cda}, Lee et al.~\cite{Lee:2015qva}, and Knapen et al.~\cite{Knapen:2021run} (with and without screening). See Sec.~\ref{subsec:compare_to_prev} for detailed comparison.}
    \label{fig:reach_compare_Q1}
\end{figure}

\begin{figure}[t]
    \centering
    \includegraphics[width=\textwidth]{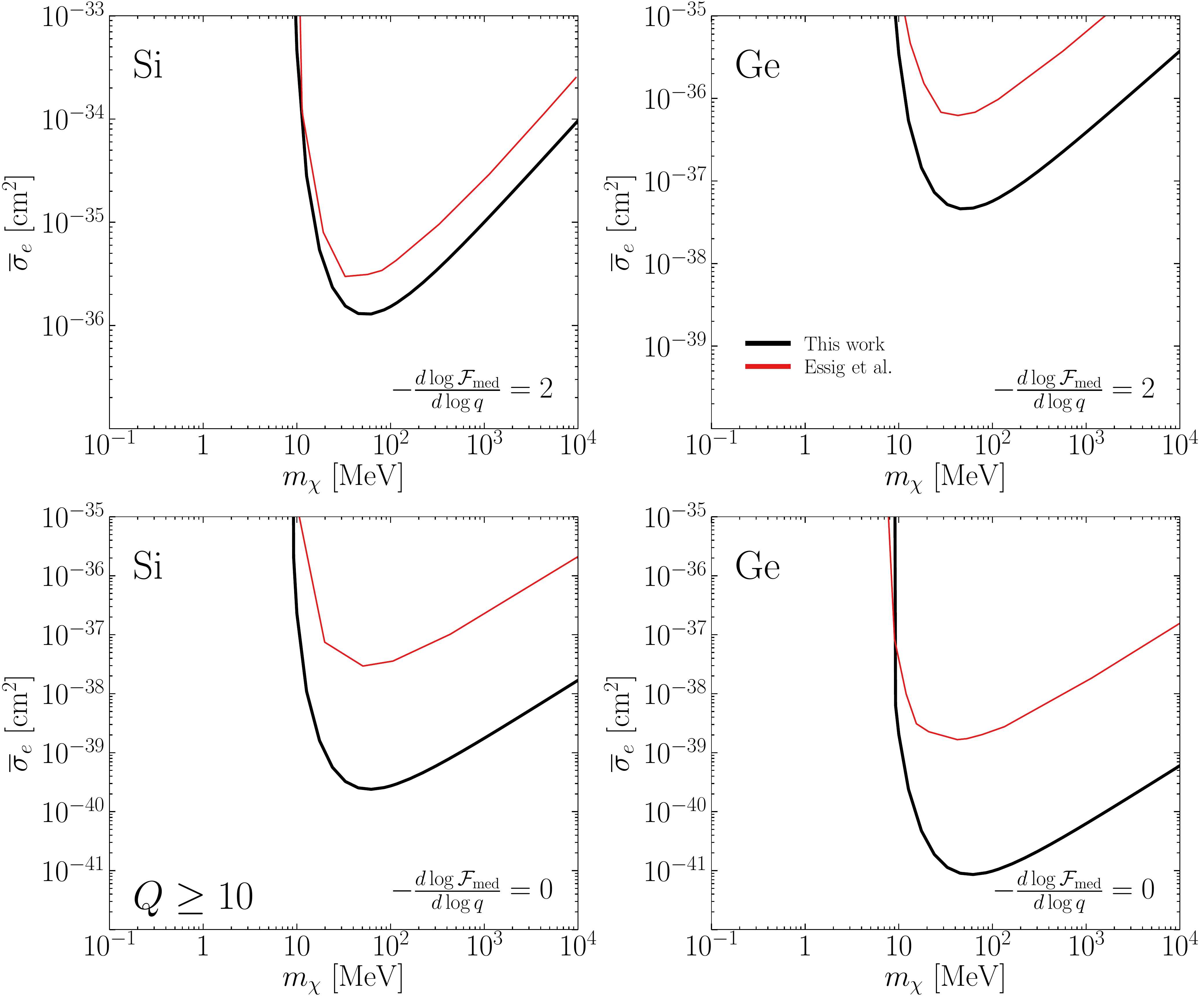}
    \caption{95\% C.L.\ exclusion reach (3 events) assuming 1 kg-year exposure, $Q \ge 10$, for light (top row) and heavy (bottom row) mediators. The results shown are from this work and Essig et al.~\cite{Essig:2015cda}. See Sec.~\ref{subsec:compare_to_prev} for detailed comparison.}
    \label{fig:reach_compare_Q10}
\end{figure}

We begin by comparing to our previous work, Ref.~\cite{Griffin:2019mvc}, shown in brown in Fig.~\ref{fig:reach_compare_Q1}. We previously restricted our analysis to the light mediator scenario, and $Q \ge 1$, which is relatively unaffected by AE reconstruction effects since the rate is peaked at small energy/momentum transfers, as seen in Fig.~\ref{fig:vc_diff_rate}. The main reason the reach here is weaker is the inclusion of in-medium screening discussed in Sec.~\ref{subsec:in-med}.  

Next we compare to Ref.~\cite{Essig:2015cda}, shown in red in Figs.~\ref{fig:reach_compare_Q1} and \ref{fig:reach_compare_Q10}. 
Those results were computed solely from valence to conduction (v$\to$c) transitions. The largest discrepancy is in the high $m_\chi$ regime scattering off a Ge target via a heavy mediator. This is due to high momentum contributions to the 3d wave functions in Ge. Ref.~\cite{Essig:2015cda} computed the 3d states with DFT without AE reconstruction, which as we saw in Fig.~\ref{fig:binned_core_wf} is crucial for recovering the dominant part of the 3d wave functions at high momentum. 
As discussed in Sec.~\ref{subsec:atomic_wf}, our modeling of 3d electrons in Ge as core states reproduces their DFT-computed wave functions up to the AE reconstruction cutoff, and provides a robust parameterization of higher momentum components.
Since the valence states in Ge also contribute an appreciable amount, the $Q \ge 1$ results in Fig.~\ref{fig:reach_compare_Q1} only differ by about an order of magnitude. 
However, the difference is more stark when going to higher $Q$ thresholds in Fig.~\ref{fig:reach_compare_Q10}, which essentially isolates the 3d electrons' contribution. 
In the low mass regime the difference is less significant, and primarily due to the inclusion of screening effects. 
Another difference that is important here is sampling of the 1BZ. Ref.~\cite{Essig:2015cda} used a uniform $6\times6\times6$ mesh with extra 27 points chosen by hand close to the center of the 1BZ, whereas here (as well as in Ref.~\cite{Griffin:2019mvc}) we use a uniform $10\times10\times10$ grid. While checking convergence we found our (unscreened) results using a $6\times6\times6$ uniform mesh were a closer match to Ref.~\cite{Essig:2015cda}; generally, increasing the number of $\vect{k}$ points reduces the rate toward convergence, {\it i.e.}\ $R_{10\times10\times10} < R_{9\times9\times9} < R_{8\times8\times8}$. This can be seen more directly in the difference between the brown and red lines in the light mediator scenario (as both are computed without screening), and it affects Ge more than Si, as is expected due to the smaller band gap and greater dispersions of nearby bands requiring denser $\vect{k}$ point sampling for convergence.

Ref.~\cite{Lee:2015qva} also computed DM-electron scattering rates in semiconductors, focusing on Ge. The approach taken in that paper was to semi-analytically model the Ge wave functions with the core wave functions (with the same set of RHF STO wave function coefficients tabulated in Ref.~\cite{Bunge:1993jsz}) and treat the final states as free with a Fermi factor, analogous to the core to free calculation performed here. As we can see from Fig.~\ref{fig:reach_compare_Q1}, while for most of the mass range and mediators the estimates are too optimistic due to incorrect modeling of the valence and conduction states, in the high mass region with a heavy mediator (bottom-right panel), where 3d states dominate, their estimates are in good agreement with ours presented here, as expected. 

Finally, we discuss the comparison with the most recent work, Ref.~\cite{Knapen:2021run}, which was limited to valence to conduction transitions. To show the effect of screening, we show their projected reach with (purple) and without (green) screening in Fig.~\ref{fig:reach_compare_Q1}. Again the largest discrepancy is in the heavy mediator scenario with a Ge target, primarily due to the neglect of the 3d states in Ref.~\cite{Knapen:2021run}. When these are not important, {\it i.e.}\ the low mass regime or a light mediator, we generally find good agreement, with our reach being a bit stronger. Notably this does not seem due to a mis-model of the dielectric, since the effect of screening relative to our previous results, Ref.~\cite{Griffin:2019mvc}, is consistent with their result. We also find that screening has a smaller effect at high masses in the heavy mediator scenario for Si. These small differences are harder to disentangle since they could be due to: 1) different xc-functionals used (PBE and HSE vs.\ TB09); 2) local field effects which are only partially included here since we assume the screening factor is isotropic; 3) the plane wave expansion parameter, $E_\text{cut}$, taken to be $500$\,eV without AE reconstruction in Ref.~\cite{Knapen:2021run}, vs.\ 1\,keV, AE corrected to 2 keV taken here; 4) DM velocity distribution parameters, studied in detail in Ref.~\cite{Radick:2020qip}, for which Ref.~\cite{Knapen:2021run} assumed $v_\text{esc} = 500$ km$/$s as opposed to $v_\text{esc} = 600$ km$/$s chosen here; and 5) Ref.~\cite{Knapen:2021run} took a directionally averaged dielectric, whereas here we only assume isotropy in the screening factor but not the matrix element itself.

\section{Conclusions}
\label{sec:conclusions}

Dark matter-electron scattering in dielectric crystal targets, especially semiconductors like Si and Ge, are at the forefront of DM direct detection experiments. 
It is therefore imperative to have accurate theoretical predictions for the excitation rates. In this work, we extended the scattering rate calculation in several key aspects. 
Much of the focus of previous calculations has been on transitions from valence to conduction bands just across the band gap, which will be accessible to near-future experiments. 
We performed state-of-the-art DFT calculations for these states, and highlighted the importance of all-electron reconstruction which has been neglected in most previous works. 
Along with this, we extended the transition rate calculation by explicitly including the contributions from core electrons and additional states more than 60\,eV above the band gap using analytic approximations. 

We updated the projected reach with our new calculation and found important differences compared to previous results. 
In particular, we found that in the heavy mediator scenario, 3d electrons in Ge give a dominant contribution to the detection rate for DM heavier than about 30\,MeV. 
Also, the rate can be significantly higher than predicted previously for higher experimental thresholds. 
This is exciting because new DM parameter space will be within reach even before detectors reach the single electron ionization threshold. 

We also release a beta version of \texttt{EXCEED-DM} (available \href{https://github.com/tanner-trickle/EXCEED-DM}{here}~\cite{tanner_trickle_2021_4747696}) that implements our DM-electron scattering calculation for general crystal targets, and make the electronic wave function data for Si and Ge~\cite{Trickle2021}, as well as the \texttt{EXCEED-DM} output~\cite{Trickle2021a}, publicly available so our present analysis can be reproduced.
We have previously used \texttt{EXCEED-DM} for a target comparison study~\cite{Griffin:2019mvc}, and to study the daily modulation signals that can arise in anisotropic materials~\cite{Trickle:2019nya}. 
The generality of \texttt{EXCEED-DM} means that potential applications are vast. 
It can be used to compute detection rates for other target materials (assuming DFT calculations of valence and conduction states are available), and can also be adapted to include additional DM interactions such as in an effective field theory framework similar to the study of atomic ionizations in Ref.~\cite{Catena:2019gfa} (see Ref.~\cite{Catena:2021qsr} for a recent effort in this direction).
For momentum-suppressed effective operators, a full calculation in our framework is even more important, as the effects of all-electron reconstruction and core states (overlooked in Ref.~\cite{Catena:2021qsr}) are generally amplified. 
Moreover, the differential information that can be obtained from our program facilitates further studies including realistic backgrounds.
Details of \texttt{EXCEED-DM} and additional example calculations will be presented in an upcoming publication.

\acknowledgments
We are grateful to Kyle Bystrom for assistance with \texttt{pawpyseed}, and thank Alex Ganose, Thomas Harrelson, Andrea Mitridate, Michele Papucci and Tien-Tien Yu for helpful discussions. 
We also thank Rouven Essig and Adrian Soto for early correspondence about \texttt{QEdark}.
This material is based upon work supported by the U.S.\ Department of Energy, Office of Science, Office of High Energy Physics, under Award No.~DE-SC0021431 (TT, ZZ, KZ), by a Simons Investigator Award (KZ) and the Quantum Information Science Enabled Discovery (QuantISED) for High Energy Physics (KA2401032).  
This research used resources of the National Energy Research Scientific Computing Center (NERSC), a U.S.\ Department of Energy Office of Science User Facility located at Lawrence Berkeley National Laboratory, operated under Contract No.~DE-AC02-05CH11231.
Some of the computations presented here were conducted on the Caltech High Performance Cluster, partially supported by a grant from the Gordon and Betty Moore Foundation. 
Work at the Molecular Foundry was supported by the Office of Science, Office of Basic Energy Sciences, of the U.S.\ Department of Energy under Contract No.~DE-AC02-05CH11231.

\bibliographystyle{utphys}
\bibliography{biblio}

\providecommand{\href}[2]{#2}\begingroup\raggedright\begin{thebibliography}{10}

\bibitem{Essig:2011nj}
R.~Essig, J.~Mardon, and T.~Volansky, ``{Direct Detection of Sub-GeV Dark
  Matter},'' \href{http://dx.doi.org/10.1103/PhysRevD.85.076007}{{\em Phys.
  Rev.} {\bf D85} (2012)  076007},
\href{http://arxiv.org/abs/1108.5383}{{\tt arXiv:1108.5383 [hep-ph]}}.

\bibitem{Graham:2012su}
P.~W. Graham, D.~E. Kaplan, S.~Rajendran, and M.~T. Walters, ``{Semiconductor
  Probes of Light Dark Matter},''
  \href{http://dx.doi.org/10.1016/j.dark.2012.09.001}{{\em Phys. Dark Univ.}
  {\bf 1} (2012)  32--49},
\href{http://arxiv.org/abs/1203.2531}{{\tt arXiv:1203.2531 [hep-ph]}}.

\bibitem{Lee:2015qva}
S.~K. Lee, M.~Lisanti, S.~Mishra-Sharma, and B.~R. Safdi, ``{Modulation Effects
  in Dark Matter-Electron Scattering Experiments},''
  \href{http://dx.doi.org/10.1103/PhysRevD.92.083517}{{\em Phys. Rev. D} {\bf
  92} (2015) no.~8, 083517}, \href{http://arxiv.org/abs/1508.07361}{{\tt
  arXiv:1508.07361 [hep-ph]}}.

\bibitem{Essig:2017kqs}
R.~Essig, T.~Volansky, and T.-T. Yu,
  \href{http://dx.doi.org/10.1103/physrevd.96.043017}{``New constraints and
  prospects for sub-{GeV} dark matter scattering off electrons in xenon,''{\em
  Physical Review D} {\bf 96} (aug, 2017)  043017},
\href{http://arxiv.org/abs/1703.00910}{{\tt arXiv:1703.00910 [hep-ph]}}.

\bibitem{Catena:2019gfa}
R.~Catena, T.~Emken, N.~A. Spaldin, and W.~Tarantino, ``{Atomic responses to
  general dark matter-electron interactions},''
  \href{http://dx.doi.org/10.1103/PhysRevResearch.2.033195}{{\em Phys. Rev.
  Res.} {\bf 2} (2020) no.~3, 033195},
  \href{http://arxiv.org/abs/1912.08204}{{\tt arXiv:1912.08204 [hep-ph]}}.

\bibitem{Agnes:2018oej}
{\bf DarkSide} Collaboration, P.~Agnes {\em et al.}, ``{Constraints on Sub-GeV
  Dark-Matter\textendash{}Electron Scattering from the DarkSide-50
  Experiment},'' \href{http://dx.doi.org/10.1103/PhysRevLett.121.111303}{{\em
  Phys. Rev. Lett.} {\bf 121} (2018) no.~11, 111303},
  \href{http://arxiv.org/abs/1802.06998}{{\tt arXiv:1802.06998 [astro-ph.CO]}}.

\bibitem{Aprile:2019xxb}
{\bf XENON} Collaboration, E.~Aprile {\em et al.}, ``{Light Dark Matter Search
  with Ionization Signals in XENON1T},''
  \href{http://dx.doi.org/10.1103/PhysRevLett.123.251801}{{\em Phys. Rev.
  Lett.} {\bf 123} (2019) no.~25, 251801},
  \href{http://arxiv.org/abs/1907.11485}{{\tt arXiv:1907.11485 [hep-ex]}}.

\bibitem{Aprile:2020tmw}
{\bf XENON} Collaboration, E.~Aprile {\em et al.}, ``{Excess electronic recoil
  events in XENON1T},''
  \href{http://dx.doi.org/10.1103/PhysRevD.102.072004}{{\em Phys. Rev. D} {\bf
  102} (2020) no.~7, 072004}, \href{http://arxiv.org/abs/2006.09721}{{\tt
  arXiv:2006.09721 [hep-ex]}}.

\bibitem{Essig:2012yx}
R.~Essig, A.~Manalaysay, J.~Mardon, P.~Sorensen, and T.~Volansky, ``{First
  Direct Detection Limits on sub-GeV Dark Matter from XENON10},''
  \href{http://dx.doi.org/10.1103/PhysRevLett.109.021301}{{\em Phys. Rev.
  Lett.} {\bf 109} (2012)  021301},
\href{http://arxiv.org/abs/1206.2644}{{\tt arXiv:1206.2644 [astro-ph.CO]}}.

\bibitem{Essig:2015cda}
R.~Essig, M.~Fernandez-Serra, J.~Mardon, A.~Soto, T.~Volansky, and T.-T. Yu,
  ``{Direct Detection of sub-GeV Dark Matter with Semiconductor Targets},''
  \href{http://dx.doi.org/10.1007/JHEP05(2016)046}{{\em JHEP} {\bf 05} (2016)
  046}, \href{http://arxiv.org/abs/1509.01598}{{\tt arXiv:1509.01598
  [hep-ph]}}.

\bibitem{Derenzo:2016fse}
S.~Derenzo, R.~Essig, A.~Massari, A.~Soto, and T.-T. Yu, ``{Direct Detection of
  sub-GeV Dark Matter with Scintillating Targets},''
  \href{http://dx.doi.org/10.1103/PhysRevD.96.016026}{{\em Phys. Rev. D} {\bf
  96} (2017) no.~1, 016026}, \href{http://arxiv.org/abs/1607.01009}{{\tt
  arXiv:1607.01009 [hep-ph]}}.

\bibitem{Hochberg:2016sqx}
Y.~Hochberg, T.~Lin, and K.~M. Zurek, ``{Absorption of light dark matter in
  semiconductors},'' \href{http://dx.doi.org/10.1103/PhysRevD.95.023013}{{\em
  Phys. Rev.} {\bf D95} (2017) no.~2, 023013},
\href{http://arxiv.org/abs/1608.01994}{{\tt arXiv:1608.01994 [hep-ph]}}.

\bibitem{Bloch:2016sjj}
I.~M. Bloch, R.~Essig, K.~Tobioka, T.~Volansky, and T.-T. Yu, ``{Searching for
  Dark Absorption with Direct Detection Experiments},''
  \href{http://dx.doi.org/10.1007/JHEP06(2017)087}{{\em JHEP} {\bf 06} (2017)
  087},
\href{http://arxiv.org/abs/1608.02123}{{\tt arXiv:1608.02123 [hep-ph]}}.

\bibitem{Kurinsky:2019pgb}
N.~A. Kurinsky, T.~C. Yu, Y.~Hochberg, and B.~Cabrera, ``{Diamond Detectors for
  Direct Detection of Sub-GeV Dark Matter},''
\href{http://arxiv.org/abs/1901.07569}{{\tt arXiv:1901.07569 [hep-ex]}}.

\bibitem{Trickle:2019nya}
T.~Trickle, Z.~Zhang, K.~M. Zurek, K.~Inzani, and S.~Griffin, ``{Multi-Channel
  Direct Detection of Light Dark Matter: Theoretical Framework},''
  \href{http://dx.doi.org/10.1007/JHEP03(2020)036}{{\em JHEP} {\bf 03} (2020)
  036}, \href{http://arxiv.org/abs/1910.08092}{{\tt arXiv:1910.08092
  [hep-ph]}}.

\bibitem{Griffin:2019mvc}
S.~M. Griffin, K.~Inzani, T.~Trickle, Z.~Zhang, and K.~M. Zurek,
  ``{Multichannel direct detection of light dark matter: Target comparison},''
  \href{http://dx.doi.org/10.1103/PhysRevD.101.055004}{{\em Phys. Rev. D} {\bf
  101} (2020) no.~5, 055004}, \href{http://arxiv.org/abs/1910.10716}{{\tt
  arXiv:1910.10716 [hep-ph]}}.

\bibitem{Griffin:2020lgd}
S.~M. Griffin, Y.~Hochberg, K.~Inzani, N.~Kurinsky, T.~Lin, and T.~Chin,
  ``{Silicon carbide detectors for sub-GeV dark matter},''
  \href{http://dx.doi.org/10.1103/PhysRevD.103.075002}{{\em Phys. Rev. D} {\bf
  103} (2021) no.~7, 075002}, \href{http://arxiv.org/abs/2008.08560}{{\tt
  arXiv:2008.08560 [hep-ph]}}.

\bibitem{Du:2020ldo}
P.~Du, D.~Egana-Ugrinovic, R.~Essig, and M.~Sholapurkar, ``{Sources of
  Low-Energy Events in Low-Threshold Dark Matter Detectors},''
  \href{http://arxiv.org/abs/2011.13939}{{\tt arXiv:2011.13939 [hep-ph]}}.

\bibitem{Hochberg:2015pha}
Y.~Hochberg, Y.~Zhao, and K.~M. Zurek, ``{Superconducting Detectors for
  Superlight Dark Matter},''
  \href{http://dx.doi.org/10.1103/PhysRevLett.116.011301}{{\em Phys. Rev.
  Lett.} {\bf 116} (2016) no.~1, 011301},
\href{http://arxiv.org/abs/1504.07237}{{\tt arXiv:1504.07237 [hep-ph]}}.

\bibitem{Hochberg:2015fth}
Y.~Hochberg, M.~Pyle, Y.~Zhao, and K.~M. Zurek, ``{Detecting Superlight Dark
  Matter with Fermi-Degenerate Materials},''
  \href{http://dx.doi.org/10.1007/JHEP08(2016)057}{{\em JHEP} {\bf 08} (2016)
  057},
\href{http://arxiv.org/abs/1512.04533}{{\tt arXiv:1512.04533 [hep-ph]}}.

\bibitem{Hochberg:2016ajh}
Y.~Hochberg, T.~Lin, and K.~M. Zurek, ``{Detecting Ultralight Bosonic Dark
  Matter via Absorption in Superconductors},''
  \href{http://dx.doi.org/10.1103/PhysRevD.94.015019}{{\em Phys. Rev.} {\bf
  D94} (2016) no.~1, 015019},
\href{http://arxiv.org/abs/1604.06800}{{\tt arXiv:1604.06800 [hep-ph]}}.

\bibitem{Hochberg:2017wce}
Y.~Hochberg, Y.~Kahn, M.~Lisanti, K.~M. Zurek, A.~G. Grushin, R.~Ilan, S.~M.
  Griffin, Z.-F. Liu, S.~F. Weber, and J.~B. Neaton, ``{Detection of sub-MeV
  Dark Matter with Three-Dimensional Dirac Materials},''
  \href{http://dx.doi.org/10.1103/PhysRevD.97.015004}{{\em Phys. Rev. D} {\bf
  97} (2018) no.~1, 015004}, \href{http://arxiv.org/abs/1708.08929}{{\tt
  arXiv:1708.08929 [hep-ph]}}.

\bibitem{Coskuner:2019odd}
A.~Coskuner, A.~Mitridate, A.~Olivares, and K.~M. Zurek, ``{Directional Dark
  Matter Detection in Anisotropic Dirac Materials},''
\href{http://arxiv.org/abs/1909.09170}{{\tt arXiv:1909.09170 [hep-ph]}}.

\bibitem{Geilhufe:2019ndy}
R.~M. Geilhufe, F.~Kahlhoefer, and M.~W. Winkler, ``{Dirac Materials for
  Sub-MeV Dark Matter Detection: New Targets and Improved Formalism},''
\href{http://arxiv.org/abs/1910.02091}{{\tt arXiv:1910.02091 [hep-ph]}}.

\bibitem{inzani2021prediction}
K.~Inzani, A.~Faghaninia, and S.~M. Griffin, ``Prediction of tunable spin-orbit
  gapped materials for dark matter detection,'' {\em Physical Review Research}
  {\bf 3} (2021) no.~1, 013069.

\bibitem{deMelloNeto:2015mca}
{\bf DAMIC} Collaboration, J.~R.~T. de~Mello~Neto {\em et al.}, ``{The DAMIC
  dark matter experiment},'' \href{http://dx.doi.org/10.22323/1.236.1221}{{\em
  PoS} {\bf ICRC2015} (2016)  1221},
  \href{http://arxiv.org/abs/1510.02126}{{\tt arXiv:1510.02126
  [physics.ins-det]}}.

\bibitem{Aguilar-Arevalo:2019wdi}
{\bf DAMIC} Collaboration, A.~Aguilar-Arevalo {\em et al.}, ``{Constraints on
  Light Dark Matter Particles Interacting with Electrons from DAMIC at
  SNOLAB},'' \href{http://dx.doi.org/10.1103/PhysRevLett.123.181802}{{\em Phys.
  Rev. Lett.} {\bf 123} (2019) no.~18, 181802},
  \href{http://arxiv.org/abs/1907.12628}{{\tt arXiv:1907.12628 [astro-ph.CO]}}.

\bibitem{Settimo:2020cbq}
{\bf DAMIC, DAMIC-M} Collaboration, M.~Settimo, ``{Search for low-mass dark
  matter with the DAMIC experiment},'' in {\em {16th Rencontres du Vietnam}:
  {Theory meeting experiment: Particle Astrophysics and Cosmology}}.
\newblock 4, 2020.
\newblock \href{http://arxiv.org/abs/2003.09497}{{\tt arXiv:2003.09497
  [hep-ex]}}.

\bibitem{Tiffenberg:2017aac}
{\bf SENSEI} Collaboration, J.~Tiffenberg, M.~Sofo-Haro, A.~Drlica-Wagner,
  R.~Essig, Y.~Guardincerri, S.~Holland, T.~Volansky, and T.-T. Yu,
  ``{Single-electron and single-photon sensitivity with a silicon Skipper
  CCD},'' \href{http://dx.doi.org/10.1103/PhysRevLett.119.131802}{{\em Phys.
  Rev. Lett.} {\bf 119} (2017) no.~13, 131802},
  \href{http://arxiv.org/abs/1706.00028}{{\tt arXiv:1706.00028
  [physics.ins-det]}}.

\bibitem{Crisler:2018gci}
{\bf SENSEI} Collaboration, M.~Crisler, R.~Essig, J.~Estrada, G.~Fernandez,
  J.~Tiffenberg, M.~Sofo~haro, T.~Volansky, and T.-T. Yu, ``{SENSEI: First
  Direct-Detection Constraints on sub-GeV Dark Matter from a Surface Run},''
  \href{http://dx.doi.org/10.1103/PhysRevLett.121.061803}{{\em Phys. Rev.
  Lett.} {\bf 121} (2018) no.~6, 061803},
  \href{http://arxiv.org/abs/1804.00088}{{\tt arXiv:1804.00088 [hep-ex]}}.

\bibitem{Abramoff:2019dfb}
{\bf SENSEI} Collaboration, O.~Abramoff {\em et al.}, ``{SENSEI:
  Direct-Detection Constraints on Sub-GeV Dark Matter from a Shallow
  Underground Run Using a Prototype Skipper-CCD},''
  \href{http://dx.doi.org/10.1103/PhysRevLett.122.161801}{{\em Phys. Rev.
  Lett.} {\bf 122} (2019) no.~16, 161801},
  \href{http://arxiv.org/abs/1901.10478}{{\tt arXiv:1901.10478 [hep-ex]}}.

\bibitem{Barak:2020fql}
{\bf SENSEI} Collaboration, L.~Barak {\em et al.}, ``{SENSEI: Direct-Detection
  Results on sub-GeV Dark Matter from a New Skipper-CCD},''
  \href{http://dx.doi.org/10.1103/PhysRevLett.125.171802}{{\em Phys. Rev.
  Lett.} {\bf 125} (2020) no.~17, 171802},
  \href{http://arxiv.org/abs/2004.11378}{{\tt arXiv:2004.11378 [astro-ph.CO]}}.

\bibitem{Agnese:2014aze}
{\bf SuperCDMS} Collaboration, R.~Agnese {\em et al.}, ``{Search for Low-Mass
  Weakly Interacting Massive Particles with SuperCDMS},''
  \href{http://dx.doi.org/10.1103/PhysRevLett.112.241302}{{\em Phys. Rev.
  Lett.} {\bf 112} (2014) no.~24, 241302},
\href{http://arxiv.org/abs/1402.7137}{{\tt arXiv:1402.7137 [hep-ex]}}.

\bibitem{Agnese:2015nto}
{\bf SuperCDMS} Collaboration, R.~Agnese {\em et al.}, ``{New Results from the
  Search for Low-Mass Weakly Interacting Massive Particles with the CDMS Low
  Ionization Threshold Experiment},''
  \href{http://dx.doi.org/10.1103/PhysRevLett.116.071301}{{\em Phys. Rev.
  Lett.} {\bf 116} (2016) no.~7, 071301},
  \href{http://arxiv.org/abs/1509.02448}{{\tt arXiv:1509.02448 [astro-ph.CO]}}.

\bibitem{Agnese:2016cpb}
{\bf SuperCDMS} Collaboration, R.~Agnese {\em et al.},
  \href{http://dx.doi.org/10.1103/PhysRevD.95.082002}{``{Projected Sensitivity
  of the SuperCDMS SNOLAB experiment},''{\em Phys. Rev.} {\bf D95} (Apr, 2017)
  082002}, \href{http://arxiv.org/abs/1610.00006}{{\tt arXiv:1610.00006
  [physics.ins-det]}}.
\url{https://link.aps.org/doi/10.1103/PhysRevD.95.082002}.

\bibitem{Agnese:2017jvy}
{\bf SuperCDMS} Collaboration, R.~Agnese {\em et al.}, ``{Low-mass dark matter
  search with CDMSlite},''
  \href{http://dx.doi.org/10.1103/PhysRevD.97.022002}{{\em Phys. Rev. D} {\bf
  97} (2018) no.~2, 022002}, \href{http://arxiv.org/abs/1707.01632}{{\tt
  arXiv:1707.01632 [astro-ph.CO]}}.

\bibitem{Agnese:2018col}
{\bf SuperCDMS} Collaboration, R.~Agnese {\em et al.},
  \href{http://dx.doi.org/10.1103/physrevlett.121.051301}{``First dark matter
  constraints from a {SuperCDMS} single-charge sensitive detector,''{\em
  Physical Review Letters} {\bf 121} (aug, 2018)  051301},
  \href{http://arxiv.org/abs/1804.10697}{{\tt arXiv:1804.10697 [hep-ex]}}.
  [Erratum: Phys.Rev.Lett. 122, 069901 (2019)].

\bibitem{Agnese:2018gze}
{\bf SuperCDMS} Collaboration, R.~Agnese {\em et al.}, ``{Search for Low-Mass
  Dark Matter with CDMSlite Using a Profile Likelihood Fit},''
  \href{http://dx.doi.org/10.1103/PhysRevD.99.062001}{{\em Phys. Rev. D} {\bf
  99} (2019) no.~6, 062001}, \href{http://arxiv.org/abs/1808.09098}{{\tt
  arXiv:1808.09098 [astro-ph.CO]}}.

\bibitem{Amaral:2020ryn}
{\bf SuperCDMS} Collaboration, D.~W. Amaral {\em et al.}, ``{Constraints on
  low-mass, relic dark matter candidates from a surface-operated SuperCDMS
  single-charge sensitive detector},''
  \href{http://dx.doi.org/10.1103/PhysRevD.102.091101}{{\em Phys. Rev. D} {\bf
  102} (2020) no.~9, 091101}, \href{http://arxiv.org/abs/2005.14067}{{\tt
  arXiv:2005.14067 [hep-ex]}}.

\bibitem{Armengaud:2018cuy}
{\bf EDELWEISS} Collaboration, E.~Armengaud {\em et al.}, ``{Searches for
  electron interactions induced by new physics in the EDELWEISS-III Germanium
  bolometers},'' \href{http://dx.doi.org/10.1103/PhysRevD.98.082004}{{\em Phys.
  Rev. D} {\bf 98} (2018) no.~8, 082004},
  \href{http://arxiv.org/abs/1808.02340}{{\tt arXiv:1808.02340 [hep-ex]}}.

\bibitem{Armengaud:2019kfj}
{\bf EDELWEISS} Collaboration, E.~Armengaud {\em et al.}, ``{Searching for
  low-mass dark matter particles with a massive Ge bolometer operated
  above-ground},'' \href{http://dx.doi.org/10.1103/PhysRevD.99.082003}{{\em
  Phys. Rev. D} {\bf 99} (2019) no.~8, 082003},
  \href{http://arxiv.org/abs/1901.03588}{{\tt arXiv:1901.03588 [astro-ph.GA]}}.

\bibitem{Arnaud:2020svb}
{\bf EDELWEISS} Collaboration, Q.~Arnaud {\em et al.}, ``{First germanium-based
  constraints on sub-MeV Dark Matter with the EDELWEISS experiment},''
  \href{http://dx.doi.org/10.1103/PhysRevLett.125.141301}{{\em Phys. Rev.
  Lett.} {\bf 125} (2020) no.~14, 141301},
  \href{http://arxiv.org/abs/2003.01046}{{\tt arXiv:2003.01046 [astro-ph.GA]}}.

\bibitem{Bunge:1993jsz}
C.~Bunge, J.~Barrientos, and A.~Bunge, ``{Roothaan-Hartree-Fock Ground-State
  Atomic Wave Functions: Slater-Type Orbital Expansions and Expectation Values
  for Z = 2-54},'' \href{http://dx.doi.org/10.1006/adnd.1993.1003}{{\em Atom.
  Data Nucl. Data Tabl.} {\bf 53} (1993)  113--162}.

\bibitem{QE-2009}
P.~Giannozzi {\em et al.}, ``Quantum espresso: a modular and open-source
  software project for quantum simulations of materials,'' {\em Journal of
  Physics: Condensed Matter} {\bf 21} (2009) no.~39, 395502 (19pp).
  \url{http://www.quantum-espresso.org}.

\bibitem{QE-2017}
P.~Giannozzi {\em et al.}, ``Advanced capabilities for materials modelling with
  quantum espresso,'' {\em Journal of Physics: Condensed Matter} {\bf 29}
  (2017) no.~46, 465901.
  \url{http://stacks.iop.org/0953-8984/29/i=46/a=465901}.

\bibitem{doi:10.1063/5.0005082}
P.~Giannozzi {\em et al.}, ``Quantum espresso toward the exascale,''
  \href{http://dx.doi.org/10.1063/5.0005082}{{\em The Journal of Chemical
  Physics} {\bf 152} (2020) no.~15, 154105},
  \href{http://arxiv.org/abs/https://doi.org/10.1063/5.0005082}{{\tt
  https://doi.org/10.1063/5.0005082}}. \url{https://doi.org/10.1063/5.0005082}.

\bibitem{Hochberg:2021pkt}
Y.~Hochberg, Y.~Kahn, N.~Kurinsky, B.~V. Lehmann, T.~C. Yu, and K.~K. Berggren,
  ``{Determining Dark Matter-Electron Scattering Rates from the Dielectric
  Function},'' \href{http://arxiv.org/abs/2101.08263}{{\tt arXiv:2101.08263
  [hep-ph]}}.

\bibitem{Knapen:2021run}
S.~Knapen, J.~Kozaczuk, and T.~Lin, ``{Dark matter-electron scattering in
  dielectrics},'' \href{http://arxiv.org/abs/2101.08275}{{\tt arXiv:2101.08275
  [hep-ph]}}.

\bibitem{Knapen:2021bwg}
S.~Knapen, J.~Kozaczuk, and T.~Lin, ``{DarkELF: A python package for dark
  matter scattering in dielectric targets},''
  \href{http://arxiv.org/abs/2104.12786}{{\tt arXiv:2104.12786 [hep-ph]}}.

\bibitem{Liang:2018bdb}
Z.-L. Liang, L.~Zhang, P.~Zhang, and F.~Zheng, ``{The wavefunction
  reconstruction effects in calculation of DM-induced electronic transition in
  semiconductor targets},''
  \href{http://dx.doi.org/10.1007/JHEP01(2019)149}{{\em JHEP} {\bf 01} (2019)
  149}, \href{http://arxiv.org/abs/1810.13394}{{\tt arXiv:1810.13394
  [cond-mat.mtrl-sci]}}.

\bibitem{Griffin:2018bjn}
S.~Griffin, S.~Knapen, T.~Lin, and K.~M. Zurek, ``{Directional Detection of
  Light Dark Matter with Polar Materials},''
  \href{http://dx.doi.org/10.1103/PhysRevD.98.115034}{{\em Phys. Rev. D} {\bf
  98} (2018) no.~11, 115034}, \href{http://arxiv.org/abs/1807.10291}{{\tt
  arXiv:1807.10291 [hep-ph]}}.

\bibitem{Coskuner:2021qxo}
A.~Coskuner, T.~Trickle, Z.~Zhang, and K.~M. Zurek, ``{Directional
  Detectability of Dark Matter With Single Phonon Excitations: Target
  Comparison},'' \href{http://arxiv.org/abs/2102.09567}{{\tt arXiv:2102.09567
  [hep-ph]}}.

\bibitem{tanner_trickle_2021_4747696}
T.~Trickle,
  \href{http://dx.doi.org/10.5281/zenodo.4747696}{``{tanner-trickle/EXCEED-DM:
  EXCEED-DM-v0.1.0},''}
\newblock Zenodo, May, 2021.
\newblock \url{https://doi.org/10.5281/zenodo.4747696}.

\bibitem{Trickle2021}
T.~Trickle, \href{http://dx.doi.org/10.5281/ZENODO.4735777}{``{EXCEED-DM:
  DFT-computed electronic wave functions for Si and Ge},''}
\newblock Zenodo, 2021.

\bibitem{Trickle2021a}
T.~Trickle, \href{http://dx.doi.org/10.5281/ZENODO.4737654}{``{EXCEED-DM:
  Calculated Scattering Rate Data for Si and Ge},''}
\newblock Zenodo, 2021.

\bibitem{Hohenberg:1964zz}
P.~Hohenberg and W.~Kohn, ``{Inhomogeneous Electron Gas},''
  \href{http://dx.doi.org/10.1103/PhysRev.136.B864}{{\em Phys. Rev.} {\bf 136}
  (1964)  B864--B871}.

\bibitem{martin2020electronic}
R.~M. Martin, {\em Electronic structure: basic theory and practical methods}.
\newblock Cambridge university press, 2020.

\bibitem{Kohn:1965zzb}
W.~Kohn and L.~J. Sham, ``{Self-Consistent Equations Including Exchange and
  Correlation Effects},''
  \href{http://dx.doi.org/10.1103/PhysRev.140.A1133}{{\em Phys. Rev.} {\bf 140}
  (1965)  A1133--A1138}.

\bibitem{Perdew1996}
J.~P. Perdew, K.~Burke, and M.~Ernzerhof, ``{Generalized Gradient Approximation
  Made Simple},'' \href{http://dx.doi.org/10.1103/PhysRevLett.77.3865}{{\em
  Physical Review Letters} {\bf 77} (1996) no.~18, 3865--3868}.
  \url{http://link.aps.org/doi/10.1103/PhysRevLett.77.3865}.

\bibitem{Heyd2003a}
J.~Heyd, G.~E. Scuseria, and M.~Ernzerhof, ``{Hybrid functionals based on a
  screened Coulomb potential},''
  \href{http://dx.doi.org/10.1063/1.1564060}{{\em Journal of Chemical Physics}
  {\bf 118} (2003) no.~18, 8207--8215}.

\bibitem{Heyd2006}
J.~Heyd, G.~E. Scuseria, and M.~Ernzerhof, ``{Erratum: Hybrid functionals based
  on a screened Coulomb potential (Journal of Chemical Physics (2003) 118
  (8207))},'' {\em Journal of Chemical Physics} {\bf 124} (2006) no.~21, .
  \url{https://doi.org/10.1063/1.2204597}.

\bibitem{Blo}
P.~Bl{\"{o}}chl, ``{Projector augmented-wave method},''
  \href{http://dx.doi.org/https://doi.org/10.1103/PhysRevB.50.17953}{{\em
  Physical Review B} {\bf 50} (1994) no.~24, 17953--17979}.

\bibitem{Kresse1999}
G.~Kresse and D.~Joubert, ``{From ultrasoft pseudopotentials to the projector
  augmented-wave method},''
  \href{http://dx.doi.org/https://doi.org/10.1103/PhysRevB.59.1758}{{\em
  Physical Review B} {\bf 59} (1999)  1758--1775}.

\bibitem{Kresse1993}
G.~Kresse and J.~Hafner, ``{Ab initio molecular dynamics for liquid metals},''
  \href{http://dx.doi.org/https://doi.org/10.1103/PhysRevB.47.558}{{\em
  Physical Review B} {\bf 47} (1993)  558--561}.

\bibitem{Kresse1994}
G.~Kresse and J.~Hafner, ``{Ab initio molecular-dynamics simulation of the
  liquid-metal–amorphous-semiconductor transition in germanium},''
  \href{http://dx.doi.org/10.1103/PhysRevB.49.14251}{{\em Physical Review B}
  {\bf 49} (1994) no.~20, 14251--14269}.
  \url{http://link.aps.org/doi/10.1103/PhysRevB.49.14251}.

\bibitem{Kresse1996}
G.~Kresse and J.~Furthm{\"{u}}ller, ``{Efficiency of ab initio total energy
  calculations for metals and semiconductors using a plane-wave basis set},''
  \href{http://dx.doi.org/https://doi.org/10.1016/0927-0256(96)00008-0}{{\em
  Computational Materials Science} {\bf 6} (1996)  15--50}.

\bibitem{Kresse1996a}
G.~Kresse and J.~Furthm{\"{u}}ller, ``{Efficient iterative schemes for ab
  initio total-energy calculations using a plane-wave basis set},''
  \href{http://dx.doi.org/10.1103/PhysRevB.54.11169}{{\em Physical Review B}
  {\bf 54} (1996)  11169--11186}.
  \url{http://link.aps.org/doi/10.1103/PhysRevB.54.11169}.

\bibitem{rostgaard2009projector}
C.~Rostgaard, ``The projector augmented-wave method,''
  \href{http://arxiv.org/abs/0910.1921}{{\tt arXiv:0910.1921
  [cond-mat.mtrl-sci]}}.

\bibitem{pawpy}
K.~{Bystrom}, D.~{Broberg}, S.~{Dwaraknath}, K.~A. {Persson}, and M.~{Asta},
  ``{Pawpyseed: Perturbation-extrapolation band shifting corrections for point
  defect calculations},''{\em arXiv e-prints} (Apr, 2019)  arXiv:1904.11572,
  \href{http://arxiv.org/abs/1904.11572}{{\tt arXiv:1904.11572
  [cond-mat.mtrl-sci]}}.

\bibitem{E.U.Condon1935}
G.~H.~S. E.~U.~Condon, {\em The Theory of Atomic Spectra}.
\newblock Cambridge University Press, January, 1935.
\newblock
  \url{https://www.ebook.de/de/product/4287067/e_u_condon_george_h_shortley_the_theory_of_atomic_spectra.html}.

\bibitem{Belkic1989}
D.~Belki{\'{c}} and H.~S. Taylor,
  \href{http://dx.doi.org/10.1088/0031-8949/39/2/004}{``A unified formula for
  the fourier transform of slater-type orbitals,''{\em Physica Scripta} {\bf
  39} (feb, 1989)  226--229}.

\bibitem{Cappellini1993}
G.~Cappellini, R.~D. Sole, L.~Reining, and F.~Bechstedt,
  \href{http://dx.doi.org/10.1103/physrevb.47.9892}{``Model dielectric function
  for semiconductors,''{\em Physical Review B} {\bf 47} (apr, 1993)
  9892--9895}.

\bibitem{Radick:2020qip}
A.~Radick, A.-M. Taki, and T.-T. Yu, ``{Dependence of Dark Matter - Electron
  Scattering on the Galactic Dark Matter Velocity Distribution},''
  \href{http://dx.doi.org/10.1088/1475-7516/2021/02/004}{{\em JCAP} {\bf 02}
  (2021)  004}, \href{http://arxiv.org/abs/2011.02493}{{\tt arXiv:2011.02493
  [hep-ph]}}.

\bibitem{Catena:2021qsr}
R.~Catena, T.~Emken, M.~Matas, N.~A. Spaldin, and E.~Urdshals, ``{Crystal
  responses to general dark matter-electron interactions},''
  \href{http://arxiv.org/abs/2105.02233}{{\tt arXiv:2105.02233 [hep-ph]}}.

\end{thebibliography}\endgroup

\end{document}